\documentclass[lettersize,journal]{IEEEtran}

\usepackage{amsmath,amsfonts}
\usepackage{algorithmic}
\usepackage{algorithm}
\usepackage{array}
\usepackage{textcomp}
\usepackage{stfloats}
\usepackage{url}
\usepackage{verbatim}
\usepackage{graphicx}
\usepackage{amssymb} % \triangleq
\usepackage{amsfonts}
\usepackage{amsmath} % align
\usepackage{physics} % 长竖线 
\usepackage{algorithmic}  % 算法框架
\usepackage{algorithm} 
\usepackage{graphicx}
\usepackage{color} % 字体颜色
\usepackage{bm}
\usepackage{epstopdf} 
\usepackage{subfigure}
\usepackage{booktabs}  % table
\usepackage[dvipsnames]{xcolor} % 更多字体颜色
\usepackage{cite} % 多项引用
\usepackage{balance}
\usepackage{setspace}  % 改变行距
\usepackage{amsthm}
\usepackage{amsmath} 
\usepackage{hyperref}
\usepackage{balance}

\allowdisplaybreaks[4]

\theoremstyle{note} % 样式设置definition为正体

\newtheorem{ppn}{Proposition}

\newtheorem*{pf}{Proof}  

\newcommand{\ppnref}[1]{\textbf{Proposition \ref{#1}}}

\definecolor{DeepPurple}{RGB}{160, 32, 240}

\newcommand{\figref}[1]{Fig. \ref{#1}}

\DeclareMathAlphabet{\mathsfit}{\encodingdefault}{\sfdefault}{m}{sl}
\SetMathAlphabet{\mathsfit}{bold}{\encodingdefault}{\sfdefault}{bx}{n}
\newcommand{\tens}[1]{\bm{\mathsfit{#1}}}

\def\tB{{\tens{B}}}
\def\tC{{\tens{C}}}
\def\tD{{\tens{D}}}

\def\tF{{\tens{F}}}
\def\tG{{\tens{G}}}

\def\tO{{\tens{O}}}

\def\tQ{{\tens{Q}}}

\def\tX{{\tens{X}}}

\begin{document}

% 调用参考文献作者缩写规则
\bstctlcite{IEEEexample:BSTcontrol}
    
% \title{Rethinking Symbol-Level Precoding: Can It Match Linear Precoding Complexity?}

\title{Unlocking Symbol-Level Precoding Efficiency Through Tensor Equivariant Neural Network}

% 标题可以自己想一下，首先是张量等变性，然后是高效，然后是符号级预编码。\textbf{符号级预编码还有其他称呼，看看现在流行的是哪一种。}
% 	可以参考庄嘉林师弟的标题“Extract the Best, Discard the Rest: CSI Feedback with Offline Large AI Models”。
    
\author{Jinshuo Zhang,  \textit{Graduate Student Member},  \textit{IEEE},  Yafei Wang,  \textit{Graduate Student Member},  \textit{IEEE},\\ Xinping Yi, \textit{Member},  \textit{IEEE},  Wenjin Wang, \textit{Member},  \textit{IEEE},  Shi Jin,  \textit{Fellow}, \textit{IEEE}, \\ Symeon Chatzinotas, \textit{Fellow}, \textit{IEEE}, Bj{\"o}rn Ottersten, \textit{Fellow}, \textit{IEEE}
% <-this % stops a space
% \thanks{Manuscript received xxx; revised xxx. This work was supported in part by the National Key Research and Development Program of China under Grant 2023YFB2904703; in part by the National Natural Science Foundation of China under Grant 62341110 and 62371122. The associate editor coordinating the review of this article and approving it for publication was xxx. \textit{(Corresponding author: Wenjin Wang.)}}
\thanks{Jinshuo Zhang, Yafei Wang, and Wenjin Wang are with the National Mobile Communications Research Laboratory, Southeast University, Nanjing 210096, China, and also with Purple Mountain Laboratories, Nanjing 211100, China (e-mail: zhangjs@seu.edu.cn; wangyf@seu.edu.cn; wangwj@seu.edu.cn).}
\thanks{Xinping Yi and Shi Jin are with the National Mobile Communications Research Laboratory, Southeast University, Nanjing 210096, China (e-mail: xyi@seu.edu.cn; jinshi@seu.edu.cn).}
\thanks{Symeon Chatzinotas and Bj{\"o}rn Ottersten are with the Interdisciplinary Centre for Security, Reliability and Trust (SnT), University of Luxembourg (E-mails: \{symeon.chatzinotas, bjorn.ottersten\}@uni.lu).}}

	\markboth{}%
	{Shell \MakeLowercase{\textit{et al.}}: A Sample Article Using IEEEtran.cls for IEEE Journals}
	
	% 页尾
	
	\maketitle

        \begin{abstract}
        Although symbol-level precoding (SLP) based on constructive interference (CI) exploitation offers performance gains, its high complexity remains a bottleneck. This paper addresses this challenge with an end-to-end deep learning (DL) framework with low inference complexity that leverages the structure of the optimal SLP solution in the closed-form and its inherent tensor equivariance (TE), where TE denotes that a permutation of the input induces the corresponding permutation of the output.
        Building upon the computationally efficient model-based formulations, as well as their known closed-form solutions, we analyze their relationship with linear precoding (LP) and investigate the corresponding optimality condition.
        We then construct a mapping from the problem formulation to the solution and prove its TE, based on which the designed networks reveal a specific parameter-sharing pattern that delivers low computational complexity and strong generalization.
        Leveraging these, we propose the backbone of the framework with an attention-based TE module, achieving linear computational complexity.
        Furthermore, we demonstrate that such a framework is also applicable to imperfect CSI scenarios, where we design a TE-based network to map the CSI, statistics, and symbols to auxiliary variables.
        Simulation results show that the proposed framework captures substantial performance gains of optimal SLP, while achieving an approximately 80-times speedup over conventional methods and maintaining strong generalization across user numbers and symbol block lengths.
	\end{abstract}

	\begin{IEEEkeywords}
		Symbol-level precoding, deep learning, tensor equivariance, imperfect channel state information.
	\end{IEEEkeywords}

	% \fi
	\IEEEpeerreviewmaketitle

	\section{Introduction}
    % 内容是错的。思路如下
    % 1. 预编码很重要
    % 2. 主要分为线性和非线性。
    % 3. 线性计算复杂度较低，但难以实现最优。
    % 3. 非线性预编码拥有更好的性能潜力，但计算复杂度高。其中SLP是一种开发有利干扰的预编码方法，它xxxxx
    \IEEEPARstart{I}{n} multi-input multi-output (MIMO) systems, precoding is a crucial technique for eliminating inter-user interference and enhancing system capacity\cite{lu2014overview,WANG2025}.
    % Precoding strategies are broadly categorized into linear precoding (LP) and nonlinear precoding. 
    Conventional linear precoding (LP), such as zero-forcing (ZF) and minimum mean square error (MMSE) precoding \cite{1468466}, has low computational complexity but fails to achieve optimality.
    % In contrast, nonlinear precoding approaches achieve superior performance but incur significantly higher computational complexity. Among them, symbol-level precoding (SLP) is a nonlinear technique that leverages both channel state information (CSI) and instantaneous transmit symbols to optimize performance on a per-symbol basis, thereby exhibiting superior performance potential 
    In contrast, symbol-level precoding (SLP) leverages both CSI and instantaneous transmit symbols to optimize performance on a per-symbol basis, thereby exhibiting superior performance potential while incurring higher computational complexity \cite{8359237,Li2021,4801492,9035662,5605266,7103338,7417066,7942010,8575164,8466792,8299553,8477154,8815429,wang2024symbol,shao2020minimum, mohammad2021unsupervised, 8647896, hegde2019interference, 9025051,7042789,7472286,8374931}.

    % Unlike conventional precoding techniques that aim to suppress multiuser interference (MUI), 
    SLP achieves performance gains by effectively leveraging constructive interference (CI), which shifts the received signal constellation points away from their decision boundaries.
    % distinguishes MUI into constructive interference (CI) and destructive interference (DI), achieving performance gains by effectively leveraging CI. 
    % For instance, the strategy of preserving CI while eliminating destructive interference (DI) was explored in \cite{4801492}, whereas \cite{5605266} investigated rotating DI to become CI. 
    % Based on CI constraints, numerous classical precoding methods have been extended to the symbol level.
    % Based on CI constraints, the symbol-level power minimization (PM) problem seeks to minimize transmit power subject to given signal-to-interference-plus-noise ratio (SINR) requirements \cite{7103338,7417066,7942010,8299553,8477154,7042789,8575164}. The symbol-level SINR balancing problem focuses on maximizing the minimum user SINR under transmit power constraints \cite{7103338,8466792,Li2021,8477154,7472286}. 
    Based on CI constraints, the symbol-level power minimization (PM) problem \cite{7103338,7417066,7942010,8299553,8477154,7042789,8575164} and signal-to-interference-plus-noise ratio (SINR) balancing problem \cite{7103338,8466792,Li2021,8477154,7472286} have been extensively studied.
    The minimization of symbol error rate (SER) was investigated in \cite{8374931,wang2024symbol,shao2020minimum}. Notably, the authors of \cite{9910472} proposed a simple and generic CI region (CIR), based on which they presented a symbol-level MMSE precoding solution.
    % and proved that the symbol-level ZF precoding is its special case in the high signal-to-noise ratio (SNR) regime.
    % 不要用转折。你这个论文不是写鲁棒预编码。你要说  准确的CSI是预编码设计的基础。
    % 不要写死，说一定不可能完美CSI，那么你的完美CSI场景就应该删除。实际系统中，基站侧的CSI可能因为xxx等原因存在一定程度的误差。如果后续这段篇幅还是不算太长，这段可以和前面一段合并。
    Moreover, available CSI may contain errors, making the modeling and estimation of imperfect channels an active research area \cite{9416909,hou2024tensor,10804143,hou2024joint,zhuang2025extract},
    % Since accurate CSI serves as the foundation for precoding, 
    Consequently, the development of robust SLP techniques has garnered significant research interest in recent years \cite{7103338, mohammad2021unsupervised, 8647896, hegde2019interference, 9025051}. In particular, leveraging the \textit{a posteriori} channel model \cite{8694866}, the work in \cite{confPaper,wang2024robust} investigated robust SLP designs optimizing for SINR and MMSE criteria under imperfect CSI, achieving promising performance.

    Despite their superior performance, SLP algorithms suffer from inherently high computational complexity due to the dependence on the instantaneous transmit symbols to all users, which limits their practical deployment.
    % 这里你要说，
    % 为了降低计算复杂度，SLP的等价设计也在不断革新。它的优化问题从最初的SOCP\cite{8466792}，到后续的QP\cite{masouros2014vector}再到NNLS\cite{8815429,9910472}，对应的最优算法复杂度不断下降。然而，它仍然面临着优化问题所存在的迭代计算复杂度高的问题。为此，有一些低计算复杂度的设计。例如文献\cite{8466792,Li2021}针对QP问题提出了一种efficient iterative algorithm，可以通过每个迭代中的闭式解obtains the optimal QP solution within only a few iterations. 文献\cite{8465957}提出了对PSK constellations 的NNLS问题的最优算法的近似算法，具有闭式计算的良好性质。
    % 这里文献的介绍要有方法，要从远离我们论文的文献一步步介绍到我们所使用的基线，告诉他们这个是最相关的，也是目前最好的，这就是我们为什么没对比其它方法的原因。
    %随着SLP等价formulation的持续性地研究，其求解的计算复杂度也随之不断降低。
    Research on equivalent SLP formulations has progressively reduced computational complexity, with the problem evolving from a second-order cone programming (SOCP) \cite{8466792} to a quadratic programming (QP) \cite{8466792, Li2021} and finally to a non-negative least squares (NNLS) formulation \cite{ 8815429, 9910472}, each step yielding more efficient algorithms.
    However, these methods still face challenges related to the high computational cost of iterative optimization compared to LP. To address this, several low-complexity designs have been proposed. For the QP formulation,  an unsupervised learning-based framework was proposed in \cite{9685567}.
    % that trains deep neural networks (DNNs) by unfolding an interior point method with a proximal logarithmic barrier function. 
    Regarding the CI-constrained PM problem, the work in \cite{8465957} proposed an approximation algorithm for the optimal solution of its NNLS formulation, which benefits from closed-form computation and demonstrates excellent performance.

    However, to the best of our knowledge, existing low-complexity schemes have not fully explored the NNLS formulation and its inherent properties to reduce computational complexity.
    % Although some closed-form NNLS solutions exist \cite{8465957,8647428}, they are restricted to specific modulations and scenarios, lacking general applicability. 
    Furthermore, although several low-complexity approaches have investigated approximate closed-form solutions \cite{8465957,8647428}, they are typically restricted to specific constellations under perfect CSI. A general method applicable to multi-level QAM and PSK under both perfect and imperfect CSI remains to be explored.
    Meanwhile, permutation equivariance/invariance (PE/PI), and their high-dimensional extension, termed tensor equivariance (TE) \cite{11049893}, have inspired dedicated TE neural networks (TENN). These networks have proven remarkably successful in LP design, detection, and soft demodulation \cite{11049893,wang2024soft,pratik2020re,wang2025statistical}, suggesting their potential in the design of SLP.
    % This leads to the central research question, 这个问题不是只在我们这几页论文里才有的
    % 我们能否通过开发其固有的张量等变性质以设计更低计算复杂度的符号级预编码？
    This leads to the central research question: \textit{How can its inherent tensor equivariance properties be exploited in designing an end-to-end learning-to-optimize solution with lower computational complexity?} This work aims to answer this question. By leveraging the NNLS formulation and its inherent TE, we develop a unified deep learning (DL)-based SLP framework supporting various modulation schemes under perfect CSI and also extending to imperfect CSI scenarios. In summary, the main contributions of this paper are as follows:
    \begin{itemize}
        \item Building upon the established NNLS formulation of typical SLP problems under SINR balancing and MMSE criteria and their known closed-form solutions, we analyze their relationship to LP and investigate the corresponding Karush-Kuhn-Tucker (KKT) conditions. Furthermore, we define a mapping from the information embedded in the KKT conditions to the optimal perturbation factors that adjust the symbols within the CIR, and then analyze the TE of this mapping, including multidimensional equivariance (MDE) and high-order equivariance (HOE).
		\item To leverage the properties of the formulated mapping, we propose a linear-complexity attention-based MDE (AMDE) module that achieves a strong representation power through a novel TE attention mechanism. Building on AMDE, we develop the SLPN network, which exploits the TE and NNLS formulation of the SLP problem, providing linear computational complexity, low parameter count, and inherent generalization for input sizes.
        The network was trained on datasets from diverse channel environments, enabling its deployment across different channel realizations.
        To further enhance performance, we propose a post-net refinement method which applies a lightweight computational step after SLPN to scale the obtained perturbation factors.
        %%%% 方法适用于多种调制方式
        By integrating these components, we construct an SLP framework that embodies these advantages, demonstrating applicability across diverse SLP schemes and constellations.
		\item For scenarios with potentially imperfect CSI, we extend the proposed SLP framework. Specifically,
        % based on the \textit{a posteriori} channel model, 
        we design a low-complexity network RSLPN to implement the MMSE robust SLP scheme, which also leverages the inherent TE. The RSLPN estimates the auxiliary variable and perturbation factor of the closed-form solution, bypassing the required iterations while preserving the advantages afforded by TE. Simulation results show that the proposed TENN-based SLP framework retain most of the performance gains of optimal SLP, while achieving an approximately 80-fold speedup over conventional methods in typical scenarios and maintaining strong generalization across both user numbers and symbol block lengths.
	\end{itemize}
    
	The rest of the article is organized as follows: The system model and CIR are built in Section \ref{Section 2}. The low complexity SLP framework based on TENN is investigated in Section \ref{perfect csi slp}, and {Section \ref{robust slp section} subsequently extends this framework to imperfect CSI scenarios.} Simulation results are provided in Section \ref{Section 6}, and Section \ref{Section 7} concludes this article.
	
	{\textit{Notation}}: $x$, ${\bf x}$, ${\bf X}$, and $\boldsymbol{\mathsf{X}}$ denote a scalar, column vector, matrix, and tensor, respectively. The operations of transpose, conjugate, conjugate transpose, and matrix inversion are denoted by $(\cdot)^T$, $(\cdot)^{*}$, $(\cdot)^H$, and $(\cdot)^{-1}$, respectively. ${\mathbb{R}}^{M\times N}$ (${\mathbb{C}}^{M\times N}$) denotes the $M\times N$ dimensional real (complex) matrix space.  diag$\{{\bf a}\}$ denotes the diagonal matrix with ${\bf a}$ on its diagonal. $\real(\cdot)$ and $\imaginary(\cdot)$ extract the real and imaginary parts of a complex quantity. $\mathcal{R}(\cdot)$ maps complex vectors/matrices to their real representations: $\mathcal{R}({\bf a})=[\real({\bf a})^T, \imaginary({\bf a})^T]^T$ and $\mathcal{R}({\bf A})=[\real({\bf A}), -\imaginary({\bf A});\imaginary({\bf A}),\real({\bf A})]$. ${\bf I}_N$ denotes the $N\times N$ identity matrix, and $\left \|\cdot\right \|_{2}$ denotes $l_2$-norm. $\otimes$ and $\odot$ are the Kronecker and Hardmard product. $[{\bf X}]_{i,j}$ and $[{\bf x}]_i$ denotes the $(i,j)$-th and $i$-th element of ${\bf X}$ and ${\bf x}$, respectively. $[{\bf X}_1,\ldots,{\bf X}_K]_{D}$ denotes the tensor formed by stacking ${\bf X}_1,\ldots,{\bf X}_K$ along the $D$-th dimension. $[\cdot]_{(D)}$ denotes the concatenation of tensors along the newly inserted $D$-th dimension.
    ${\bf x}{\succeq}{\bf 0}$ means all the elements of ${\bf x}$ are nonnegative.
    % $\in$ denotes``belongs to'', $\sim$ denotes ``be distributed as'', $\triangleq$ denotes the definition, and  
    $\mathcal{C}\mathcal{N}(\mu, \sigma^2)$ denotes circularly symmetric Gaussian distribution with expectation $\mu$ and variance $\sigma^2$. The Kronecker product of tensor and matrix is defined as $(\tX \otimes_n {\bf Y})_{[m_1,\ldots,m_{n-1},:,:,m_{n+2},\ldots,m_{N}]}=\tX_{[m_1,\ldots,m_{n-1},:,:,m_{n+2},\ldots,m_{N}]}\otimes {\bf Y}$.

\section{System Model and Symbol-Level Precoding}\label{Section 2}
\subsection{System Model}\label{system model}

    Consider a MIMO downlink transmission system, where a base station (BS) with $N_{\rm T}$ antennas serves $K$ single-antenna user equipments (UE). Each transmission time slot comprises two phases: (i) an uplink pilot phase for CSI acquisition, and (ii) a subsequent downlink data phase transmitting $L$ symbols.
    {We assume block flat fading channels, whose coefficients remain constant over a coherence interval of $L$ symbol durations. Let ${\bf h}_{k}\in{\mathbb{C}}^{N_{\rm T}\times 1}$ denote the channel vector between the BS and the $k$-th user, which is obtained during the CSI acquisition phase.}
    The signal received by the $k$-th user at the $l$-th downlink symbol is given by
	\begin{equation}
		{y}_{k}[l] = {\bf h}_{k}^T{\bf x}_{{\rm c}}[l] + n_{k}[l],\;\forall k \in {\mathcal{ K}},\forall l \in {\mathcal{ L}},
		\label{received signal 1}
	\end{equation}
        where ${\mathcal{ L}}\triangleq\left\{1, 2, ..., L\right\}$, $\mathcal{K}\triangleq\{1,2,\ldots,K\}$, ${\bf x}_{{\rm c}}[l]\in{\mathbb{C}}^{N_{\rm T}\times 1}$ denotes the precoded signal vector transmitted by the BS during the $l$-th symbol, and $n_{k}[l]$ denotes the noise at the $k$-th user, i.e., $n_{k}[l]\sim\mathcal{C}\mathcal{N}(0, \sigma^2)$.

\subsection{Constructive Interference Region} \label{sec:CIR}

\begin{figure}[t]
		\centering
		\includegraphics[width=3.2in]{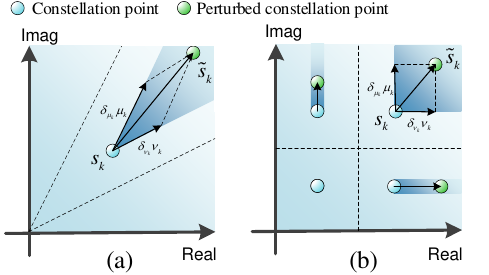}
		\caption{CIRs and their boundary vectors of (a) PSK and (b) QAM.}
		\label{CI_MMSE_CIR}
	\end{figure}

    {CI refers to interference that shifts the received signal constellation point further from the decision boundaries \cite{9910472, 8466792, Li2021}. For a specific transmit symbol, its CIR is defined as the area in the complex plane encompassing all signal points that undergo such CI. \figref{CI_MMSE_CIR} illustrates the CIR (deep blue regions), boundary vectors, and maximum likelihood (ML) decision boundaries (dashed lines) for both PSK and QAM modulations.
    The typical CIR of $s_{k}$ can be expressed as \cite{9910472}}
	\begin{align}
		\mathcal{D}_k = \left\{{{{\tilde {s}}_k}}|
		{{\tilde {s}}_k} = s_k + \delta_{\mu_k}{{\mu}}_{k}+\delta_{\nu_k}{{\nu}}_{k}
		,\ \delta_{\mu_k},\delta_{\nu_k}>0\right\},
		\label{CIR-D}
	\end{align}
    where $\mu_k$ and $\nu_k$ are two complex parameters representing the boundaries of the CIR for
    % the transmit symbol 
    $s_k$, while $\delta_{\mu_k}$ and $\delta_{\nu_k}$ are perturbation factors, shifting the transmit symbol within the CIR. For notational convenience, we define $\boldsymbol{\mu}=[{{\mu}}_{1},\cdots ,{{\mu}}_{K}]^{T}$, $\boldsymbol{\nu}=[{{\nu}}_{1},\cdots ,{{\nu}}_{K}]^{T}$, $\boldsymbol{\Lambda}_\mu=\text{diag}\{\boldsymbol{\mu}\}$, $\boldsymbol{\Lambda}_\nu=\text{diag}\{\boldsymbol{\nu}\}$, $\boldsymbol{\delta}_\mu=[\delta_{\mu_{1}},\ldots,\delta_{\mu_{K}}]^{T}$, $\boldsymbol{\delta}_\nu=[\delta_{\nu_{1}},\ldots,\delta_{\nu_{K}}]^{T}$, ${\bf s}_{{\rm{c}}} = \left[{s}_{1},{s}_{2},..., {s}_{K}\right]^T$, and $\tilde{\bf s}_{{\rm{c}}} = \left[\tilde{s}_{1},\tilde{s}_{2},..., \tilde{s}_{K}\right]^T$, then $\tilde{\mathbf{s}}_{\mathrm{c}}$ is expressed as
    \begin{align}
        \tilde{\bf s}_{\rm c}&={\bf s}_{\rm c}+{\boldsymbol{\Lambda}_\mu}{\boldsymbol{\delta}_\mu}+{\boldsymbol{\Lambda}_\nu}{\boldsymbol{\delta}_\nu}.\label{eq:cir complex form}
	\end{align}
    Let ${\bf s}=\mathcal{R}({\bf s}_{\rm c})$ and $\tilde{\bf s}=\mathcal{R}(\tilde{\bf s}_{\rm c})$ be the real-valued representations of ${\bf s}_{\rm c}$ and $\tilde{\bf s}_{\rm c}$, respectively. Define $\boldsymbol{\delta}=[\delta_{\mu_{1}},\ldots,\delta_{\mu_{K}},\delta_{\nu_{1}},\ldots,\delta_{\nu_{K}}]^{T}$ and
    \begin{align}
		\begin{split}
			&{\boldsymbol{\Lambda}} = 
			\begin{bmatrix}
				\text{diag}\{\real(\boldsymbol{\mu})\}&\text{diag}\{\real(\boldsymbol{\nu})\}\\
				\text{diag}\{\imaginary(\boldsymbol{\mu})\}&\text{diag}\{\imaginary(\boldsymbol{\nu})\}
			\end{bmatrix},
		\end{split} \label{eq:gamma define}
	\end{align}
	thus the real-valued equivalent representation of $\tilde{\bf s}_{\rm c}$ is given by
	\begin{align}
		{\tilde{\bf s}}&={\bf s}+{\boldsymbol{\Lambda}}{\boldsymbol{\delta}} \label{cir-represant}.
	\end{align}
    This representation provides an effective formulation of the CIR, establishing a foundation for analyzing SLP problems and deriving their solutions in the subsequent sections.

% \vspace{-13pt}
\subsection{Optimization-Based SLP Design}\label{sec:opt-based slp trans}

    The correct demodulation of multi-level QAM requires knowledge of the scaling factor at the receiver. Denoting the scaling factor for the $l$-th symbol by $\gamma[l]$, which can be sent by the BS or estimated at the receiver, the signal for demodulation at the $k$-th UE is expressed as\cite{wang2024symbol,8602458,li2020symbol}
	\begin{align}
		\begin{split}
			{\tilde{y}_{k}[l]} &= {y}_{k}[l]/{\gamma[l]} = {\bf h}_{k}^T{\bf x}_{{\rm c}}[l]/{\gamma[l]} + n_{k}[l]/{\gamma[l]},
		\end{split}\label{eq:receive signal for modulation}
	\end{align}
    
    \subsubsection{CIZF}
    Although the CIZF precoding is essentially a CI-constrained SINR balancing (CISB) problem, its closed-form expression resembles that of ZF, so we refer to it as CIZF precoding. Specifically, it maximizes the scaling factor while satisfying the power constraint, expressed by \cite{7103338,li2020symbol}
    \begin{align}
        \begin{aligned}
            &\max_{{\bf x}_c[l],\tilde{\bf s}_{\rm c}[l],\gamma[l]} \  \gamma[l]\\
            & \text{s.t.} \ {\bf h}_{k}^T{\bf x}_{c}[l]=\gamma[l]\cdot\tilde{s}_k[l],\ \forall k\in\mathcal{K},\\
            &\quad \ \ \tilde{s}_k[l]\in\mathcal{D}_k[l],\ \forall k\in\mathcal{K},\\
            &\quad\ \ \|{\bf x}_{\rm c}[l]\|_2^2\le P_{\rm T},\label{eq:cizf-problem-persymbol}
        \end{aligned}
    \end{align}
    where $P_{\rm T}$ denotes the transmit power. This problem can be optimally addressed by first solving a convex PM problem and then scaling the solution to meet the power constraint \cite{8815429}. The PM problem is formulated as follows
    \begin{align}
        \begin{aligned}
            &\min_{\boldsymbol{\delta}_{\mu}[l]\succeq{\bf 0},\boldsymbol{\delta}_{\nu}[l]\succeq{\bf 0}}\ \tilde{\mathbf{s}}_{{\rm c}}^H[l]{\boldsymbol{\Upsilon}_{\rm ZF}}\tilde{\mathbf{s}}_{{\rm c}}[l]\\
        &\text{s.t.} \ \tilde{\mathbf{s}}_{{\rm c}}[l]={\mathbf{s}}_{{\rm c}}[l]+\boldsymbol{\Lambda}_{\mu}[l]\boldsymbol{\delta}_{\mu}[l]+\boldsymbol{\Lambda}_{\nu}[l]\boldsymbol{\delta}_{\nu}[l],
        \end{aligned}\label{eq:cizf-problem-upslion}
    \end{align}
    where $\boldsymbol{\Upsilon}_{\rm ZF}=(\mathbf{H}\mathbf{H}^H)^{-1}$ and $\mathbf{H}=[{\bf{h}}_{1},\cdots ,{\bf{h}}_{K}]^{T}$. 
    {This problem is equal to the following NNLS problem}
    \begin{align}
        \min_{\boldsymbol{\delta}[l]\succeq{\bf 0}} \ \|\mathbf{H}^\dagger_{\rm r}\boldsymbol{\Lambda}[l]\boldsymbol{\delta}[l]+\mathbf{H}^\dagger_r{{\bf s}[l]}\|_2^2,
        \label{eq:cizf-nnls}
    \end{align}
    {where $\mathbf{H}^\dagger_r=\mathcal{R}(\mathbf{H}^\dagger)\in\mathbb{R}^{2N_{\rm T}\times 2K}$ and  $\mathbf{H}^\dagger=\mathbf{H}^H(\mathbf{H}\mathbf{H}^H)^{-1}$. The definitions of $\boldsymbol{\delta}[l]$, $\boldsymbol{\Lambda}[l]$, and ${{\bf s}[l]}$ are provided in Section \ref{sec:CIR}.}
    {This NNLS problem can be solved by the active set-based algorithm \cite{lawson1995solving} through an iterative procedure. After further considering the transmit power constraint, the structure of the optimal solution to problem \eqref{eq:cizf-problem-persymbol} is expressed as}
    \begin{align}
            {\bf x}_{{{\rm{c}}}}^\star[l]&=\gamma^\star[l]\mathbf{H}^\dagger\tilde{\mathbf{s}}_{{\rm c}}^\star[l], \
            \gamma^\star[l]=\sqrt{{P_{\mathrm{T}}}/{\|\mathbf{H}^\dagger\tilde{\mathbf{s}}_{{\rm c}}^\star[l]\|_2^2}},\label{eq:cizf no norm 2}\\
            \tilde{\mathbf{s}}_{{\rm c}}^\star[l]&={\mathbf{s}}_{{\rm c}}[l]+\boldsymbol{\Lambda}_{\mu}[l]\boldsymbol{\delta}_{\mu}[l]+\boldsymbol{\Lambda}_{\nu}[l]\boldsymbol{\delta}_{\nu}[l].\label{eq:cizf no norm 3}
    \end{align}

% \vspace{-5pt}
    \subsubsection{CIMMSE}
    The MMSE problem minimizes the mean square error (MSE) between transmit and receive symbols, and with CI constraints it is formulated as \cite{9910472}
    \begin{align}
		\begin{aligned}&\min_{{\bf x}_c[l],\tilde{\bf s}_{\rm c}[l],\gamma[l]}\mathbb{E}_{\mathbf{n}[l]}\left\{\left\|\left({\mathbf{H}}\mathbf{x}_{{\rm c}}[l]+{\mathbf{n}[l]}\right)/{\gamma[l]}-\tilde{\mathbf{s}}_{{\rm c}}[l]\right\|_2^2\right\}\\
        &\ \ \quad  \text{s.t.}\ \tilde{s}_{k}[l]\in\mathcal{D}_{k}[l],\ \forall k\in\mathcal{K},\\
        &\ \ \quad\ \quad \|{\bf x}_{{\rm c}}[l]\|_2^2\le P_{{\rm T}},\end{aligned}\label{eq:cimmse-problem}
	\end{align}
    {where $\mathbf{n}[l]=\left[{{n}}_{1}[l],\cdots ,{{n}}_{K}[l]\right]^{T}$.}
    This problem can be optimally solved by applying a scaling method to the solution of its convex power-unconstrained counterpart, which is stated as follows \cite{9910472}
    \begin{align}
        \begin{aligned}
            &\min_{\boldsymbol{\delta}_{\mu}[l]\succeq{\bf 0},\boldsymbol{\delta}_{\nu}[l]\succeq{\bf 0}}\ \tilde{\mathbf{s}}_{{\rm c}}^H[l]\boldsymbol{\Upsilon}_{\rm MMSE}\tilde{\mathbf{s}}_{{\rm c}}[l]\\
            &\text{s.t.} \ \tilde{\mathbf{s}}_{{\rm c}}[l]={\mathbf{s}}_{{\rm c}}[l]+\boldsymbol{\Lambda}_{\mu}[l]\boldsymbol{\delta}_{\mu}[l]+\boldsymbol{\Lambda}_{\nu}[l]\boldsymbol{\delta}_{\nu}[l],\label{eq:te design cimmse p1}
        \end{aligned}
    \end{align}
    where $\boldsymbol{\Upsilon}_{\rm MMSE}=\left({\mathbf{H}}{\mathbf{H}}^H+\frac{\sigma^2K}{P_{{\rm T}}}\mathbf{I}_{K}\right)^{-1}$. The above problem can be formulated as an NNLS problem
    \begin{align}
        \min_{\boldsymbol{\delta}[l]\succeq{\bf 0}} \ \|{\bf C}_{\rm u}\boldsymbol{\Lambda}[l]\boldsymbol{\delta}[l]+{\bf C}_{\rm u}{\bf s}[l]\|_2^2,\label{eq:cimmse-nnls}
    \end{align}
    where ${\bf C}_{\rm u}$ denotes the upper triangular matrix obtained from the Cholesky decomposition of $\left({\bf H}_{\rm r}{\bf H}_{\rm r}^T+\frac{\sigma^2K}{P_{\rm T}}{\bf I}_{2K}\right)^{-1}$, with ${\bf H}_{\rm r}=\mathcal{R}({\bf H})$. This NNLS problem can be solved using a similar approach as that for problem \eqref{eq:cizf-nnls}. Accounting for power constraints, the solution exhibits the following structure
    \begin{align}
			{\mathbf{x}}_{{\rm c}}^\star[l]&=\gamma^\star[l]{\mathbf{H}}^H\boldsymbol{\Upsilon}_{\rm MMSE}\tilde{\mathbf{s}}_{{\rm{c}}}^\star[l],\label{eq:cimmse no norm 1}\\
			\gamma^\star[l]&=\sqrt{{P_{{\rm T}}}/{\|{\mathbf{H}}^H\boldsymbol{\Upsilon}_{\rm MMSE}\tilde{\mathbf{s}}_{{\rm{c}}}^\star[l]\|_2^2}},\label{eq:cimmse no norm 2}\\
            \tilde{\mathbf{s}}_{{\rm c}}^\star[l]&={\mathbf{s}}_{{\rm c}}[l]+\boldsymbol{\Lambda}_{\mu}[l]\boldsymbol{\delta}_{\mu}[l]+\boldsymbol{\Lambda}_{\nu}[l]\boldsymbol{\delta}_{\nu}[l].\label{eq:cimmse no norm 3}
	\end{align}
    As established in \cite{9910472}, the optimal solution to the CIMMSE problem reduces to that of the CIZF problem in the high SNR regime (i.e., as $\sigma \rightarrow 0$).
    
    Furthermore, a block-level power reallocation scheme is introduced to maintain a uniform rescaling factor $\bar\gamma$ over the entire block, which substantially reduces the per-symbol overhead \cite{li2020symbol,wang2024soft}:
    \begin{align}
        \bar{\gamma}^\star=\sqrt{\frac{L}{\sum_{l=1}^L\frac{1}{(\gamma^\star[l])^2}}},\ \ \bar{\mathbf{x}}_{{\rm{c}}}^\star[l]=\frac{\bar{\gamma}^\star}{\gamma^\star[l]}\mathbf{x}_{{\rm{c}}}^\star[l], \ \forall l\in\mathcal{L}.\label{eq:norm gamma}
    \end{align}
    where $\bar\gamma^\star$ can be acquired at the UE either via the control channel or through estimation.
    
    It is worth noting that, for the CIZF problem, the solution obtained through the above power allocation method exactly coincides with the optimal solution of the block-level CISB problem that minimizes $\bar{\gamma}$ over a symbol block of length $L$ and subject to the total transmit energy constraint $\sum_{l=1}^{L}\|{\bf x}_{\rm c}[l]\|_2^2 \leq L P_{\rm T}$ \cite{li2020symbol}.
    % 于是the scaled signal 可以被进一步表述为
    Accordingly, the signal for demodulation in \eqref{eq:receive signal for modulation} can be reformulated as
    \begin{align}
	\begin{split}
		{\tilde{y}_{k}[l]} &= {y}_{k}[l]/{\bar\gamma} = {\bf h}_{k}^T\bar{\bf x}_{{\rm c}}[l]/{\bar\gamma} + n_{k}[l]/{\bar\gamma}.
	\end{split}
    \end{align}

\subsection{Symbol-Level Precoding vs. Linear Precoding} \label{sec:SLPvsLP}
    \subsubsection{Sources of SLP Gains}
    {By comparing the formulations of CIZF in \eqref{eq:cizf no norm 2},\eqref{eq:cizf no norm 3}, CIMMSE in \eqref{eq:cimmse no norm 1}-\eqref{eq:cimmse no norm 3}, and LP schemes such as ZF and MMSE\cite{1468466}, we can observe a key difference in the general expression between SLP and LP schemes, specifically}
    \begin{align}
        {\rm LP:}&\ {\bf x}_{\rm LP}=\gamma_{\rm LP}{\bf P}{\bf s}_{\rm c},\\
        {\rm SLP:}&\ {\bf x}_{\rm SLP}=\gamma_{\rm SLP}{\bf P}({\mathbf{s}_{\rm c}}+\boldsymbol{\Lambda}_{\mu}\boldsymbol{\delta}_{\mu}+\boldsymbol{\Lambda}_{\nu}\boldsymbol{\delta}_{\nu}),
    \end{align}
    where ${{\bf P}\in\mathbb{C}^{N_{\rm T}\times K}}$ is the precoding matrix, $\gamma_{\rm LP}$ and $\gamma_{\rm SLP}$ are the power normalization factors. 
    The key distinction of SLP from LP is its intentional perturbation of the symbols to exploit the CIR, which is the primary mechanism behind its performance gains. From a multicast perspective, these gains stem from exploiting interference constructively as an additional source of useful power \cite{7942010}. Consequently, the core challenge of SLP lies in determining the perturbation factors $\boldsymbol{\delta}_{\mu}[l]$ and $\boldsymbol{\delta}_{\nu}[l]$. With these factors obtained, the SLP problems can be solved using the closed-form solutions given by equations \eqref{eq:cizf no norm 2}, \eqref{eq:cizf no norm 3}, \eqref{eq:cimmse no norm 1}-\eqref{eq:cimmse no norm 3}, and \eqref{eq:norm gamma}. Moreover, when $N_{\rm T} \gg K$, the gain introduced by SLP vanishes under most channel conditions, with $\boldsymbol{\delta}_{\mu}[l]$ and $\boldsymbol{\delta}_{\nu}[l]$ converging to $\mathbf{0}$ \cite{wang2024robust}.

    \subsubsection{Gap in Computational Efficiency}
    {The complexity of LP mainly arises from $\bf Ps$ multiplication, which is $\mathcal{O}(KN_{\rm T})$ and can be efficiently computed in batches. Conversely, SLP incurs a higher complexity of $\mathcal{O}\bigl(KN_{\rm T}+KN_{\rm T}N_L+N_{\rm T}N_L^3\bigr)$, where $N_L$ denotes the number of main loop iterations of the active set-based algorithm. Furthermore, SLP requires distinct iterative computations for each symbol, which makes it challenging to directly implement efficient batch processing.}

\section{A Low-Complexity Symbol-Level Precoding Framework Based on Tensor Equivariance}\label{perfect csi slp}
{In this section, we propose a low-complexity framework based on TENN for solving the SLP problem. We begin by analyzing the shared formulation of the CIZF and CIMMSE problems and deriving their KKT conditions. Building on this analysis, we design a mapping from the available information of the KKT conditions into perturbation factors and subsequently analyze its inherent TE. Following this, we propose a linear-complexity attention-based MDE module, which forms the foundation for the SLP framework.}

    %基于KKT的
    \subsection{KKT-Based SLP Mapping}\label{sec:gradient mapping}
    {As analyzed in Section \ref{sec:SLPvsLP}, the key to solve the CIZF \eqref{eq:cizf-problem-persymbol} and CIMMSE \eqref{eq:cimmse-problem} problems lies in obtaining $\boldsymbol{\delta}_{\mu}[l]$ and $\boldsymbol{\delta}_{\nu}[l]$, which can be derived by solving the problems in \eqref{eq:cizf-problem-upslion} and \eqref{eq:te design cimmse p1}.
    Given the consistency in form between the two problems, we use the general notation $\boldsymbol{\Upsilon}$ to replace $\boldsymbol{\Upsilon}_{\rm ZF}$ and $\boldsymbol{\Upsilon}_{\rm MMSE}$  to facilitate a unified analysis. 
To gain further insight into solving this problem, we derive its Lagrangian as follows
    \begin{align}
        \begin{aligned}
        &\mathcal{L}({\boldsymbol{\delta}}_{\mu}[l],{\boldsymbol{\delta}}_{\nu}[l],{\boldsymbol{\lambda}}_{\mu}[l],{\boldsymbol{\lambda}}_{\nu}[l])\\&=\tilde{\mathbf{s}}_{{\rm c}}^H[l]\boldsymbol{\Upsilon}\tilde{\mathbf{s}}_{{\rm c}}[l]+\boldsymbol{\lambda}^T_{\mu}[l]\boldsymbol{\delta}_{\mu}[l]+\boldsymbol{\lambda}^T_{\nu}[l]\boldsymbol{\delta}_{\nu}[l],
        \end{aligned}
    \end{align}
    where $\boldsymbol{\lambda}_{\mu}[l]$ and $\boldsymbol{\lambda}_{\nu}[l]$ are Lagrange multiplier vectors. 
    Therefore, the KKT conditions can be expressed as
\begin{subequations}\begin{align}\nabla_{\boldsymbol{\delta}_{\mu}[l]}\mathcal{L}({\boldsymbol{\delta}}_{\mu}[l],{\boldsymbol{\delta}}_{\nu}[l],{\boldsymbol{\lambda}}_{\mu}[l],{\boldsymbol{\lambda}}_{\nu}[l])&=0,\label{eq:kkt1}\\\nabla_{\boldsymbol{\delta}_{\nu}[l]}\mathcal{L}({\boldsymbol{\delta}}_{\mu}[l],{\boldsymbol{\delta}}_{\nu}[l],{\boldsymbol{\lambda}}_{\mu}[l],{\boldsymbol{\lambda}}_{\nu}[l])&=0,\label{eq:kkt2}\\
{\boldsymbol{\delta}}_{\mu}[l]\succeq 0,\ {\boldsymbol{\lambda}}_{\mu}[l]\preceq 0,\ {\boldsymbol{\lambda}}_{\mu}^T[l]{\boldsymbol{\delta}}_{\mu}[l]&=0,\label{eq:kkt3}\\
{\boldsymbol{\delta}}_{\nu}[l]\succeq 0,\ {\boldsymbol{\lambda}}_{\nu}[l]\preceq 0,\ {\boldsymbol{\lambda}}_{\nu}^T[l]\label{eq:kkt4}{\boldsymbol{\delta}}_{\nu}[l]&=0,
    \end{align}
    \end{subequations}
    and we have
    \begin{align}
        &\nabla_{\boldsymbol{\delta}_{\mu}[l]}\mathcal{L}({\boldsymbol{\delta}}_{\mu}[l],{\boldsymbol{\delta}}_{\nu}[l],{\boldsymbol{\lambda}}_{\mu}[l],{\boldsymbol{\lambda}}_{\nu}[l]) \label{eq:gradient mu}={\boldsymbol{\lambda}}_{\mu}[l]+\\&2\real(\boldsymbol{\Lambda}_{\mu}^H[l]\boldsymbol{\Upsilon}{\mathbf{s}}_{{\rm c}}[l]\!+\!\boldsymbol{\Lambda}_{\mu}^H[l]\boldsymbol{\Upsilon}\boldsymbol{\Lambda}_{\mu}[l]\boldsymbol{\delta}_{\mu}[l]\!+\!\boldsymbol{\Lambda}_{\mu}^H[l]\boldsymbol{\Upsilon}\boldsymbol{\Lambda}_{\nu}[l]\boldsymbol{\delta}_{\nu}[l]),\notag\\
        &\nabla_{\boldsymbol{\delta}_{\nu}[l]}\mathcal{L}({\boldsymbol{\delta}}_{\mu}[l],{\boldsymbol{\delta}}_{\nu}[l],{\boldsymbol{\lambda}}_{\mu}[l],{\boldsymbol{\lambda}}_{\nu}[l])\label{eq:gradient nu}={\boldsymbol{\lambda}}_{\nu}[l]+\\&2\real(\boldsymbol{\Lambda}_{\nu}^H[l]\boldsymbol{\Upsilon}{\mathbf{s}}_{{\rm c}}[l]\!+\!\boldsymbol{\Lambda}_{\nu}^H[l]\boldsymbol{\Upsilon}\boldsymbol{\Lambda}_{\mu}[l]\boldsymbol{\delta}_{\mu}[l]\!+\!\boldsymbol{\Lambda}_{\nu}^H[l]\boldsymbol{\Upsilon}\boldsymbol{\Lambda}_{\nu}[l]\boldsymbol{\delta}_{\nu}[l]).\notag
    \end{align}
    Since problems \eqref{eq:cizf-problem-upslion} and \eqref{eq:te design cimmse p1} can be mathematically reformulated as problems \eqref{eq:cizf-nnls} and \eqref{eq:cimmse-nnls}, respectively, the aforementioned KKT conditions also hold for problems \eqref{eq:cizf-nnls} and \eqref{eq:cimmse-nnls}, differing only in their real-valued and complex-valued representations.
    We then define the bias term ${\bf B}_{\rm c}[l]$ and the coefficient term ${\tC}_{\rm c}[l]$ of the above gradient expressions, as follows
    \begin{align}
        {\bf B}_{\rm c}[l]&=\left[\boldsymbol{\Lambda}_{\mu}^H[l]\boldsymbol{\Upsilon}{\mathbf{s}}_{{\rm c}}[l],\boldsymbol{\Lambda}_{\nu}^H[l]\boldsymbol{\Upsilon}{\mathbf{s}}_{{\rm c}}[l]\right]\in\mathbb{C}^{K\times2},\label{eq:cal gradient 1}\\
        {\tC}_{\rm c}[l]&=\big[\boldsymbol{\Lambda}_{\mu}^H[l]\boldsymbol{\Upsilon}\boldsymbol{\Lambda}_{\mu}[l],\boldsymbol{\Lambda}_{\mu}^H[l]\boldsymbol{\Upsilon}\boldsymbol{\Lambda}_{\nu}[l],\notag\\&\boldsymbol{\Lambda}_{\nu}^H[l]\boldsymbol{\Upsilon}\boldsymbol{\Lambda}_{\mu}[l],\boldsymbol{\Lambda}_{\nu}^H[l]\boldsymbol{\Upsilon}\boldsymbol{\Lambda}_{\nu}[l]\big]_3\in\mathbb{C}^{K\times K\times 4}\label{eq:cal gradient 2},
    \end{align}
    where ${\bf B}_{\rm c}[l]$ and ${\tC}_{\rm c}[l]$ contain the available information from the KKT conditions. 
    We define ${\bf D}[l]=[\boldsymbol{\delta}_{\mu}^\star[l], \boldsymbol{\delta}_{\nu}^\star[l]]\in\mathbb{R}^{K\times 2}$, and by stacking all the relevant information of the $L$ symbols together, we obtain
    \begin{align}
        {\tB}_{\rm c}&=\big[{\bf B}_{\rm c}[1],{\bf B}_{\rm c}[2],\dots,{\bf B}_{\rm c}[L]\big]_{(2)}\in\mathbb{C}^{K\times L\times 2}\label{eq:cal gradient 3},\\
        {\tC}_{\rm c}&=\big[{\tC}_{\rm c}[1],{\tC}_{\rm c}[2],\dots,{\tC}_{\rm c}[L]\big]_{(3)}\in\mathbb{C}^{K\times K\times L\times 4}\label{eq:cal gradient 4},\\
        {\tD}&=\big[{\bf D}[1],{\bf D}[2],\dots,{\bf D}[L]\big]_{(2)}\in\mathbb{R}^{K\times L\times 2}\label{eq:definition of D}.
    \end{align}
    Since \eqref{eq:cizf-problem-upslion} and \eqref{eq:te design cimmse p1} are convex problems, their optimal solution $\tD^\star$ can be determined by their KKT information $\tB_{\rm c}$ and $\tC_{\rm c}$.
    We define a mapping from the available information within the KKT conditions to the optimal solution of the problem:
    \begin{align}
        G(\tB_{\rm c},\tC_{\rm c})=\tD^\star. \label{eq:the mapping}
    \end{align}
    Although the optimization problems across symbol periods are theoretically decoupled, we define a unified mapping over a block of $L$ symbols. This design aims to enable the subsequent proposed neural network to capture the underlying statistical patterns and shared structures that persist across different symbol groups. Through offline training on channel realizations in a pre-collected dataset, the neural network is expected to learn an approximation of mapping $G$, enabling deployment to unseen channel realizations. Moreover, in the subsequent parts, we further exploit properties of this mapping to achieve a significant reduction in both computational complexity and parameter count, as well as strong generalization with respect to user numbers and symbol block lengths.

    \subsection{Tensor Equivariance in SLP Design}\label{subsection TE in SLP design}
    Before analyzing the properties of the mapping $G$, we first introduce the definition of TE.
    We begin with the PE in the one-dimensional (1-D) case. Consider ${\bf d}\in\mathbb{C}^{D\times 1}$, $f: \mathbb{C}^{D\times 1}\to\mathbb{C}^{O\times 1}$, and let $\mathbb{S}_D$ denote the set of all permutations of $D$ indices. For any permutation $\pi_{D} \in \mathbb{S}_D$, the operator $\pi_{D}\circ_n$ applies $\pi_{D}$ along the $n$-th dimension\cite{zaheer2017deep}. Assuming $D=O$, the map $f$ exhibits PE if\cite{hartford2018deep,yun2019transformers,kim2022pure}
    \begin{align}
		f(\pi_D\circ_1{\bf d})=\pi_D\circ_1f({\bf d}),\ \forall\pi_D\in\mathbb{S}_{D}.
	\end{align}
    This implies that a permutation applied to the input induces a corresponding permutation in the ordering of elements in the output. Similarly, if the mapping $f$ exhibits PI, we have\cite{lee2019set}
    \begin{align}
	f(\pi_D\circ_1{\bf d})=f({\bf d}),\ \forall\pi_D\in\mathbb{S}_{D}.
    \end{align}
    This implies that permuting the indices of the input does not alter the output. \figref{MDEMDIpicture} illustrates some examples of PE and PI. When both the input and output are tensors, and PE holds independently across multiple dimensions, the property is referred to as MDE or $N$-D equivariance. Conversely, if PI holds across multiple dimensions, the property is termed multidimensional invariance (MDI) or $N$-D invariance. When the same PE is satisfied across multiple dimensions of both the input and the output, it is defined as HOE or $p$-$q$ order equivariance \cite{keriven2019universal}. Furthermore, TE serves as the collective term for these higher-dimensional extensions of equivariance/invariance properties, following the detailed definitions in \cite{11049893}.

    \begin{figure}[t]
	\centering
	\includegraphics[width=3.0in]{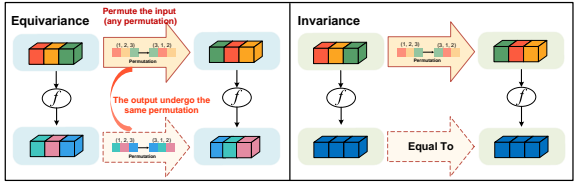}
	\caption{Some examples of equivariance and invariance.}
	\label{MDEMDIpicture}
    \end{figure}

    % \subsubsection{{Tensor Equivariance of the SLP Mapping}}
    \begin{ppn}\label{ppn: te for cizf problem}
	The mapping $G(\tB_{\rm c},\tC_{\rm c})=\tD^\star$ \eqref{eq:the mapping} defined based on problems for CIZF and CIMMSE in \eqref{eq:cizf-problem-persymbol} and \eqref{eq:cimmse-problem} exhibits the following properties
    \begin{subequations}
        \begin{align}
            G(\pi_{K}\circ_1 {\tB_{\rm c}},\pi_{K}\circ_{[1,2]}\tC_{\rm c}) &=\pi_{K}\circ_1\tD^\star,\ \forall\pi_K\in\mathbb{S}_K,\label{eq:cizf pos TE a}\\ 
            G(\pi_{L}\circ_2 {\tB_{\rm c}},\pi_{L}\circ_3\tC_{\rm c}) &=\pi_{L}\circ_2\tD^\star,\ \forall\pi_L\in\mathbb{S}_L.\label{eq:cizf pos TE b}
        \end{align}
    \end{subequations}
    \end{ppn}
    \begin{pf}
	See Appendix \ref{proof: te for cizf problem}.
    \end{pf}

    According to \ppnref{ppn: te for cizf problem}, $\tB_{\rm c}$ exhibits 2-D equivariance with respect to its first and second dimensions, and $\tC_{\rm c}$ exhibits 2-1-order equivariance with respect to its first and second dimensions and 1-D equivariance with respect to its third dimension. 

    The essence of designing neural networks for symbol-level precoding (SLP) lies in their ability to approximate the mapping $G$ effectively. In \ppnref{ppn: te for cizf problem}, we have analyzed the TE of $G$, which suggests that networks aiming to approximate $G$ would benefit from possessing similar TE. As established in \cite{11049893,pan2022permutation,zaheer2017deep}, networks exhibiting such TE inherently incorporate distinctive parameter-sharing patterns, which lead to benefits such as reduced online computational complexity and inherent generalization for input sizes. Building on this insight, we next present the design of lightweight yet efficient TE networks with excellent generalization capabilities.

\subsection{Attention-Based Multidimensional Equivariant Module}
    To achieve high performance while ensuring compliance with TE, we design an attention-based residual module, referred to as AMDE, which satisfies MDE while maintaining low online computational complexity. Since AMDE is constructed from basic TE layers \cite{11049893}, we begin by providing a brief introduction to them.

    When a fully connected layer satisfies MDE or HOE, it follows a distinct parameter-sharing pattern. We use $f_{\rm MDE}$ and $f_{\rm HOE}$ to denote the functions that obey these MDE and HOE patterns, respectively. The formal proof is provided in \cite{11049893}, where several variants of this pattern are also introduced.  In particular, the 2-D equivariant layer with input ${\bf X}\in\mathbb{R}^{M_1\times M_2\times F}$ can be expressed as \cite{11049893}
	\begin{align}
		\begin{aligned}
		    f_{\mathrm{MDE}}(\tX)&=\sum\nolimits_{\mathcal{D}\in{\mathbb{S}_2}}\bar{\tX}_{\mathcal{D}}\!\times\! {\bf W}_{\mathcal{D}}\!+\!{\bf 1}\otimes_2{\bf b}^{T}\\
        &={\tX}_\emptyset\!\times\! {\bf W}_\emptyset\!+\!\bar{\tX}_{\{1\}}\!\times\! {\bf W}_{\{1\}}\!+\!\bar{\tX}_{\{2\}}\!\times\! {\bf W}_{\{2\}}\\
        &\qquad\! +\!\bar{\tX}_{\{1,2\}}\!\times\! {\bf W}_{\{1,2\}}\!+\!{\bf 1}\otimes_2{\bf b}^{T},
		\end{aligned}
	\end{align}
	where $M_1$ and $M_2$ are its equivariant dimensions, and $F$ is the feature dimension. $\bar{\mathbf{X}}_{\mathcal{D}}$ denotes the tensor obtained by averaging ${\bf X}$ along the dimensions $\mathcal{D}$ and then broadcasting back to the original shape, and $\bar{\mathbf{X}}_\emptyset=\mathbf{X}$. The set $\mathbb{S}_2=\{\emptyset,\{1\},\{2\},\{1,2\}\}$ contains all possible combinations of the equivariant dimensions. ${\bf W}_{\mathcal{D}}\in\mathbb{R}^{F\times F_{\rm O}}$ and ${\bf b}\in\mathbb{R}^{F_{\rm O}\times 1}$ are learnable parameters, where $F_{\rm O}$ is the length of the output features. 
    Furthermore, for the MDI, we adopt the module proposed in \cite{11049893}, which utilizes a multi-head attention mechanism to capture invariance across multiple dimensions, thereby significantly outperforming traditional methods such as mean or max pooling.

    \begin{figure*}[t]
	\centering
	\includegraphics[width=0.95\textwidth]{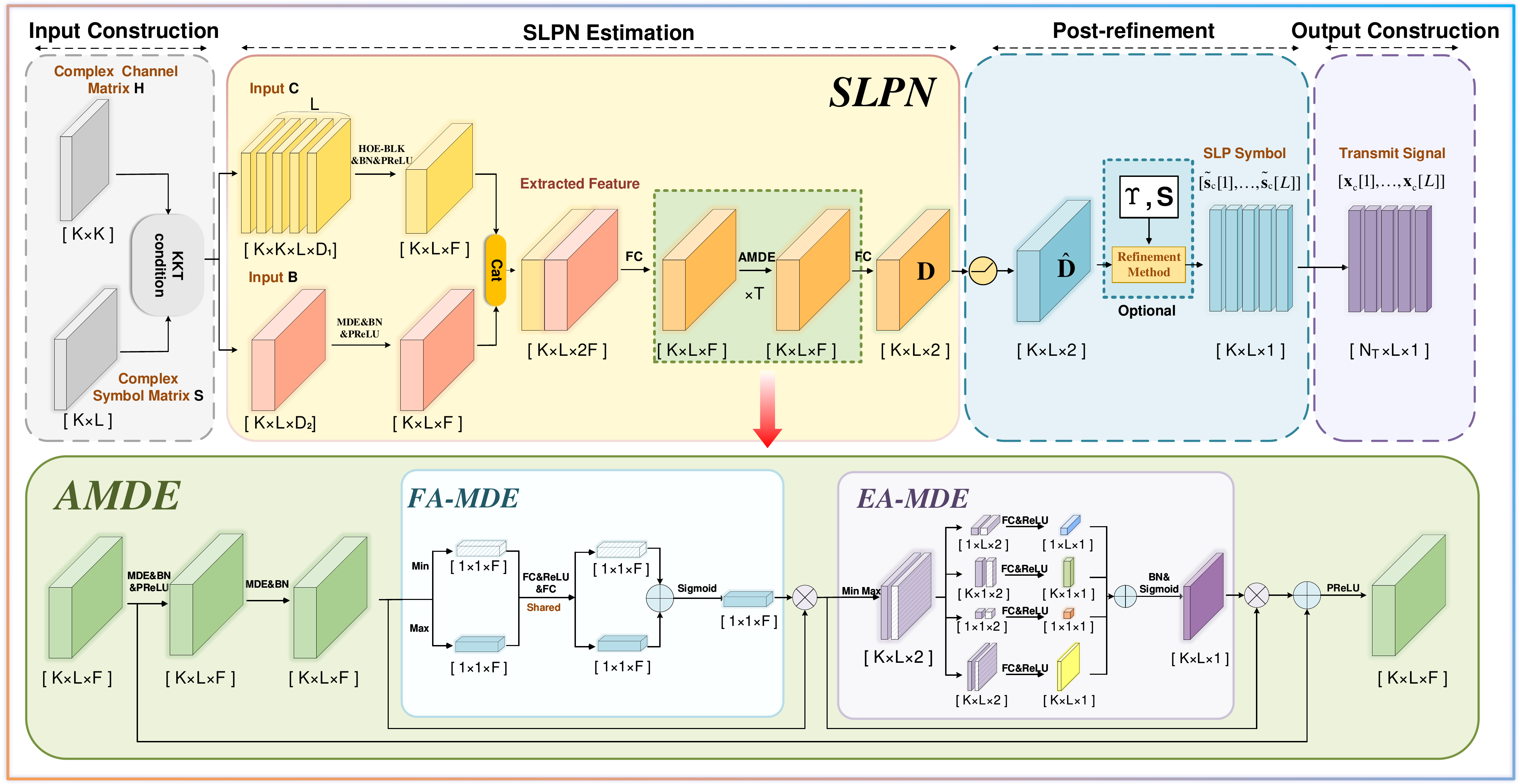}
	\caption{The overall structure of the SLP framework.}
	\label{pic:TECIZF}
    \end{figure*}

    Building upon the basic TE layers, we designed the AMDE module, with its 2-D architecture illustrated in Fig. \ref{pic:TECIZF}. The core of AMDE is a novel, lightweight decoupled attention mechanism compliant with MDE. This mechanism separately processes attention submodules along the feature and equivariant dimensions, and then serially integrates both components, following a design philosophy similar to that in \cite{Woo_2018_ECCV}. The submodules are deliberately designed to ensure strict compliance with MDE, thereby enhancing representational capacity while preserving online computational efficiency. Furthermore, the residual mechanism is integrated to alleviate training difficulties \cite{He_2016_CVPR}. 
    The formulation of AMDE is given as follows
    \begin{subequations}
        \begin{align}
        &\tX'=\text{PReLU}(\text{BN}(f_\text{MDE}(\tX))),\\
        &\tX''=\text{BN}(f_\text{MDE}(\tX')),\\
        &\tF=f_{\text{FA-MDE}}(\tX'')\odot\tX'',\\
        &\tF'=f_{\text{EA-MDE}}(\tF)\odot\tF,\\
        &\tO=\text{PReLU}(\tF'+\tX),
    \end{align}
    \end{subequations}
    where $\odot$ denotes element-wise multiplication during which the values are broadcast accordingly. BN refers to the batch normalization operation on the equivariant dimensions.

    The $f_{\text{FA-MDE}}$ represents the feature attention MDE module, which performs pooling operations along equivariant dimensions and establishes inter-feature relationships, thereby generating a feature attention map that emphasizes the most informative components within the feature dimension of the input tensor. It can be expressed as
    \begin{align}
        f_{\text{FA-MDE}}(\tX)&=\text{Sigmoid}\left(f(\tX^{\rm max}_{\mathcal{D}})+f({\tX}^{\rm mean}_{\mathcal{D}})\right),\\
        f(\tX)&=\text{FC}(\text{ReLU}(\text{FC}(\tX))),
    \end{align}
    where $\text{FC}(\cdot)$ denotes the application of a fully connected layer along the feature dimension of the input tensor, and the set $\mathcal{D}$ contains the dimensions for which pooling operations are performed. $\tX^{\rm mean}_\mathcal{D}$ and $\tX^{\rm max}_\mathcal{D}$ denote the tensors resulting from taking the mean and the maximum of $\tX$ over the dimensions in $\mathcal{D}$, respectively. The two $f(\cdot)$ share the same weights. For 2-D equivariance, $\mathcal{D}$ can be $\{1\}$, $\{2\}$, or $\{1,2\}$, and including more dimensions in $\mathcal{D}$ reduces computational cost, whereas fewer dimensions yield higher representation power. 
    
    The $f_{\text{EA-MDE}}$ is the equivariant-dimension attention MDE module that utilizes the relationship between equivariant dimensions to generate an attention map that conforms to MDE, which can be represented as
    \begin{align}
        f_{\text{EA-MDE}}(\tX)&=\text{Sigmoid}\!\left(f_{\rm RMDE}([\tX^{\rm max}_{\{n_{\rm F}\}},\tX^{\rm mean}_{\{n_{\rm F}\}}]_{n_{\rm F}})\!\right),\\
        f_{\rm RMDE}(\tX)&=\sum\nolimits_{\mathcal{D}\in{\mathbb{S}_{D}}}\text{ReLU}\left(\text{FC}(\bar{\tX}_{\mathcal{D}})\right),
    \end{align}
    where $n_{\rm F}$ is the feature dimension. Compared to the basic MDE layer, the RMDE layer applies a ReLU activation to the dimension-averaged outputs before the summation, aiming to enhance the network's nonlinear representation power.

        The proposed AMDE module is constructed from basic TE layers, which are characterized by low computational complexity \cite{11049893}. Specifically, the complexity of the module is primarily dominated by its MDE layer, leading to an overall complexity of $\mathcal{O}(2^N \bar{M} F F_{\rm O})$ \cite{11049893}, where $\bar{M} = \Pi_{n=1}^{N} M_n$ denotes the product of the lengths of $N$ equivariant dimensions. The complexity thus scales linearly with $\bar{M}$, which confirms the computationally efficient nature of the proposed module.
        Furthermore, owing to its specific parameter-sharing pattern, AMDE module has a parameter count of $\mathcal{O}(2^N F F_{\rm O})$, which is independent of the lengths of the equivariant dimensions. Similarly, the basic TE layers maintain a parameter count independent of the input sizes \cite{11049893}. Consequently, these modules and the networks built upon them exhibit inherent adaptability to varying input sizes \cite{11049893,pan2022permutation,hartford2018deep}.
    
        \begin{algorithm}[t]
		% \setstretch{1.35}
		\caption{SLP-DL Algorithm for CIZF Precoding}
		\label{A1}
		\begin{spacing}{1.2}
			\begin{algorithmic}[1]
				\STATE \textbf{Input:} ${{\bf H}}$, ${\bf S}=[{\mathbf{s}}_{{\rm c}}[1],\ldots,{\mathbf{s}}_{{\rm c}}[L]]$.
				% \\\textbf{Gradient Extraction:} 
                \STATE Get $\boldsymbol{\Upsilon}=$$(\mathbf{H}\mathbf{H}^H)^{-1}$.
                \STATE  Get $\boldsymbol{\Lambda}_{\mu}[l]$ and $\boldsymbol{\Lambda}_{\nu}[l]$ according to the definition of CIR.
                \STATE  Determine ${\tB}_{\rm c}$ and ${\tC}_{\rm c}$ according to \eqref{eq:cal gradient 1}-\eqref{eq:cal gradient 4}.
                \STATE Construct $\tB$ and $\tC$ using \eqref{eq:cal input 1},\eqref{eq:cal input 2}.
                % \\\textbf{SLPN:}
                \STATE $\tD=\text{SLPN}(\tB,\tC)$.
                % \\\textbf{Obtain the Optimized Symbol:}
                \STATE Determine $\hat{\boldsymbol{\delta}}_{\mu}[l]$,$\hat{\boldsymbol{\delta}}_{\nu}[l]$ using \eqref{eq:relu D},\eqref{eq:hatD to delta}.
                \STATE Get $\rho[l]$ using \eqref{eq:get factor alpha}.
                \STATE Get ${\tilde{\bf s}_{\rm c}[l]}$ using \eqref{eq:factor s}.
                \STATE Get ${\bf x}_{\rm c}[l]$ and $\gamma[l]$ using \eqref{eq:cizf no norm 2}.
                \STATE Get $\bar\gamma$, and $\bar{\bf x}_{{\rm c}}[l]$ using \eqref{eq:norm gamma}.
				\STATE \textbf{Output:}  $\bar{\bf x}_{{\rm c}}[l], \bar\gamma, \forall l=1,\ldots,L$.
			\end{algorithmic}
		\end{spacing}
	\end{algorithm}

\subsection{TE-Based Low-Complexity SLP Framework}\label{LC robust SINR}
    Building upon the basic TE layers and AMDE module, we propose a network for SLP named SLPN, which approximates the mapping $G(\tB_{\rm c},\tC_{\rm c})=\tD^\star$ \eqref{eq:the mapping}. This network satisfies all the required TE while maintaining low online computational complexity, and its parameter-sharing structure offers inherent generalization for input sizes. Based on this, we further develop an SLP framework, whose overall architecture is illustrated in Fig. \ref{pic:TECIZF}. Specifically, the framework consists of the following components
    
    \subsubsection{Input Construction}
    Given the channel matrix ${\bf H}\in\mathbb{C}^{N_{\rm T}\times K}$ and the transmit symbol ${\bf S}\in\mathbb{C}^{K\times L}$, we first compute the matrix $\boldsymbol{\Upsilon}$ and apply Frobenius normalization to it. The KKT information $\tB_{\rm c}$ and $\tC_{c}$  are then constructed according to \eqref{eq:cal gradient 1}–\eqref{eq:cal gradient 4}. Their real and imaginary components are then separated and concatenated along the last dimension to form the input tensors $\tC$ and $\tB$, defined as follows
    \begin{align}
        {\tC}&=[{\real (\tC_{\rm c})},{\imaginary ({\tC_{\rm c}})}]_4\in\mathbb{R}^{K\times K\times L\times D_1},\label{eq:cal input 1}\\
        {\tB}&=[{\real (\tB_{\rm c})},{\imaginary (\tB_{\rm c})}]_3\in\mathbb{R}^{K\times L\times D_2},\label{eq:cal input 2}
    \end{align}
    where $D_1=8$ and $D_2=4$.
    \subsubsection{SLPN Estimation}
    As shown in Fig. \ref{pic:TECIZF}, the SLPN first extracts feature tensors from $\tC$ and $\tB$ using the 2-1 order HOE and 2-D MDE layers, respectively. Here, the HOE-BLK sequentially performs $f_{\mathrm{HOE}}$, BN, SiLU, and FC layers. These feature tensors are subsequently combined and processed by an FC layer, reducing the feature dimension from $2F$ to $F$, and it is further refined through a stack of $T$ 2-D AMDE modules, with $\mathcal{D}$ of FA-MDE being $\{1,2\}$. Finally, a concluding FC layer is applied to reduce the feature dimension from $F$ to 2.
        \begin{align}
            \tC'&=\text{SiLU}({\rm BN}(f_{\rm HOE}(\tC)))\in\mathbb{R}^{K\times L\times F},\\
            \tC''&=\text{PReLU}({\rm BN}({\rm FC}(\tC')))\in\mathbb{R}^{K\times L\times F},\\
            \tB'&=\text{PReLU}({\rm BN}(f_{\rm MDE}(\tB)))\in\mathbb{R}^{K\times L\times F},\\
            \tF&={\rm FC}([\tC'',\tB']_3)\in\mathbb{R}^{K\times L\times F},\\
            \tD&={\rm FC}(f_{\text{AMDE}}^{\times T}(\tF))\in\mathbb{R}^{K\times L\times 2},
        \end{align}
    where $f_{\text{AMDE}}^{\times T}(\cdot)$ denotes the stacking of $T$ consecutive AMDE modules.

    \subsubsection{Post-Net Refinement Method}
    To enforce non-negativity, we apply a ReLU activation to $\tD$, yielding
    \begin{align}
        \hat{\tD}={\rm ReLU}(\tD). \label{eq:relu D}
    \end{align}
    Consequently, $\hat{\boldsymbol{\delta}}_{\mu}[l]$ and $\hat{\boldsymbol{\delta}}_{\nu}[l]$ are obtained by decomposing $\hat{\tD}$,  expressed as
    \begin{align}
        \begin{aligned}
            \hat{\boldsymbol{\delta}}_{\mu}[l]=\hat{\tD}_{[:,l,1]}, \ \hat{\boldsymbol{\delta}}_{\nu}[l]=\hat{\tD}_{[:,l,2]}.\label{eq:hatD to delta}
        \end{aligned}
    \end{align}
    An optimization factor $\rho[l]$ is introduced to scale $\hat{\boldsymbol{\delta}}_{\mu}[l]$ and $\hat{\boldsymbol{\delta}}_{\nu}[l]$, with the goal of further enhancing performance. The corresponding optimization problem is formulated as follows
    \begin{align}
        \begin{aligned}
			&\min_{\rho[l] \ge0} \sum\nolimits_{l=1}^{L} ({\mathbf{s}}_{{\rm c}}[l]+\rho[l]{\bf p}[l])^H\boldsymbol{\Upsilon}({\mathbf{s}}_{{\rm c}}[l]+\rho[l]{\bf p}[l])\\
			&\text{s.t.} \ {\bf p}[l]=\boldsymbol{\Lambda}_{\mu}[l]\hat{\boldsymbol{\delta}}_{\mu}[l]+\boldsymbol{\Lambda}_{\nu}[l]\hat{\boldsymbol{\delta}}_{\nu}[l].
		\end{aligned}
    \end{align}
    The optimal solution is given by
    \begin{align}
		\rho[l]\!=\!
        \begin{cases}
            \max\left\{0, -\frac{{\mathbf{s}}_{{\rm c}}^H[l]\boldsymbol{\Upsilon}{\bf p}[l]+{\bf p}^H[l]\boldsymbol{\Upsilon}{\mathbf{s}}_{{\rm c}}[l]}{2{\mathbf{p}}^H[l]\boldsymbol{\Upsilon}{\bf p}[l]}\right\} & \text{if}\ {\bf p}[l] \ne {\bf 0}\\
            0&\text{if}\ {\bf p}[l] = {\bf 0}.
        \end{cases}\label{eq:get factor alpha}
	\end{align}
    Finally, the transmit symbols are given by
    \begin{align}
        \tilde{\mathbf{s}}_{{\rm c}}[l]={\mathbf{s}}_{{\rm c}}[l]+\rho[l](\boldsymbol{\Lambda}_{\mu}[l]\hat{\boldsymbol{\delta}}_{\mu}[l]+\boldsymbol{\Lambda}_{\nu}[l]\hat{\boldsymbol{\delta}}_{\nu}[l]).\label{eq:factor s}
    \end{align}
    The proposed post-net refinement method is a heuristic approach that can be selectively integrated to enhance performance, depending on the specific optimization problem and training methodology.

    \subsubsection{Output Construction}
    For CIZF, the solutions $\bar{\bf x}_{\rm c}[l]$ and $\bar\gamma$ are obtained from \eqref{eq:cizf no norm 2} and \eqref{eq:norm gamma}, whereas for CIMMSE they are determined by \eqref{eq:cimmse no norm 1}, \eqref{eq:cimmse no norm 2}, and \eqref{eq:norm gamma}.

    By integrating the above framework, we propose the SLP-DL algorithm, with its CIZF version detailed in Algorithm \ref{A1}. For CIMMSE, the algorithm requires two modifications: (i) In Step 2, $\boldsymbol{\Upsilon}$ is modified to $\left({\mathbf{H}}{\mathbf{H}}^H+\frac{\sigma^2K}{P_{{\rm T}}}\mathbf{I}_{K}\right)^{-1}$; (ii) In Step 10, ${\bf x}_{\rm c}[l]$ and $\gamma[l]$ are obtained using equations \eqref{eq:cimmse no norm 1} and \eqref{eq:cimmse no norm 2}.

    We employ supervised learning, utilizing the following mean squared error loss for training
    \begin{align}
        {\rm Loss}=\frac{1}{N_{\rm sp}}\sum\nolimits_{s=1}^{N_{\rm sp}}{\rm MSE}\left({\tD}[s],\tD^\star[s]\right), \label{eq:loss mse}
    \end{align}
    where $\tD[s]$ and $\tD^\star[s]$ denote the predicted and target tensors for the $s$-th sample in the dataset, respectively, and $N_{\rm sp}$ is the total number of samples. The MSE is computed as the average of squared differences over all tensor elements.

\section{Low-Complexity Robust Symbol-Level Precoding Framework Against Channel Aging}\label{robust slp section} 
    The CSI may contain errors due to channel estimation errors \cite{hou2025tensor} and channel aging \cite{9416909}, necessitating the development of robust precoding strategies to mitigate the associated performance degradation. Conventional robust SLP approaches, which rely on optimization-based methods, often suffer from excessive computational complexity. To overcome this challenge, we first derive a closed-form robust SLP solution under the MMSE criterion, then analyze the inherent TE of the problem, and subsequently propose a tensor-equivariant and low-complexity DL framework for the efficient design of robust SLP.
    \subsection{Robust SLP Transmission}
    We utilize the \textit{a posteriori} channel model to characterize the propagation channels \cite{8694866}. Let ${\bf h}_{k}[l]\in{\mathbb{C}}^{N_{\rm T}\times 1}$ denote the channel vector between the BS and the $k$-th user during the $l$-th downlink symbol interval, and let ${\bf h}_{k}[0]$ denote the channel estimated during the uplink pilot transmission phase, then ${\bf h}_{k}[l]$ can be represented as follows \cite{confPaper}
    \begin{align}
	\begin{split}
	  {\bf h}_{k}[l] &= \alpha_{k}[l] {\bf h}_{k}[0]+ \sqrt{1-\alpha_{k}^2[l]}{\bf V}_{\rm T}^{*} ( {\bf m}_{k}\odot {\bf w}_{k}[l]).
	\end{split}
	\label{posteriori}
    \end{align}
    In this formulation, the temporal evolution of the channel is modeled as a first-order Markov process. The term $\alpha_{k}[l]$ represents the time correlation coefficient of the $l$-th symbol interval for user $k$, which are characterized using Jakes’ autocorrelation model; The matrix ${\bf V}_{\rm T}\in{\mathbb{C}}^{N_{\rm T}\times N_FN_{\rm T}}$ comprises a partial discrete Fourier transform (DFT) matrix, with $N_F\in{\mathbb{N}}^{+}$ being the fine factor that enhances model accuracy; The $ {\bf m}_{k}\in{\mathbb{R}}^{N_FN_{\rm T} \times 1}$ is a sparse and nonnegative vector, whose elements remain constant for a long time; The elements of ${\bf w}_{k}[l]\in{\mathbb{C}}^{N_FN_{\rm T} \times 1}$ are independent and identically distributed (i.i.d.), following $\mathcal{C}\mathcal{N}(0, 1)$.

    By defining $\bar{\bf h}_{k}[l]=\alpha_{k}[l]{\bf h}_{k}[0]$, substituting \eqref{posteriori} into \eqref{received signal 1} and subsequently simplifying the resulting expression, we obtain \cite{wang2024robust}
    \begin{align}
        y_{k}[l]=\bar{\bf h}_{k}^T[l]{\bf x}_{{\rm c}}[l]+\bar{n}_{k}[l],
    \end{align}
    where $\bar{n}_{k}[l]=(\beta_{k}[l]{\bf V}_{k}{\bf x}_{{\rm c}}[l])^T$${\bf w}_{k}[l]$$+n_{k}[l]$, with $\beta_{k}[l]=\sqrt{1-\alpha_{k}^2[l]}$; ${\mathbf{V}}_{k}=[\mathbf{m}_{k}\odot\mathbf{v}_{1}\cdots\mathbf{m}_{k}\odot\mathbf{v}_{N_{\rm T}}]$, with ${\bf v}_i$ being the $i$-th row of ${\bf V}_{\rm T}$. It can be verified that $\bar{n}_{k}[l]\sim\mathcal{CN}(0,\beta_{k}^2[l]\|{\mathbf{V}}_k\mathbf{x}_{{\rm c}}[l]\|_2^2+\sigma^2)$ \cite{wang2024robust}.
    Thus, the signal for demodulation can be expressed as
	\begin{align}
		\begin{split}
			{\tilde{y}_{k}[l]} &= {y}_{k}[l]/{\gamma_{k}[l]} = \bar{\bf h}_{k}^T[l]{\bf x}_{{\rm c}}[l]/{\gamma_{k}[l]} + \bar{n}_{k}[l]/{\gamma_{k}[l]},
		\end{split}
	\end{align}
    where $\gamma_{k}[l]$ is a user-specific rescaling factor of the SLP, which enhances robustness against imperfect CSI but is only applicable to PSK modulation.
    
    The robust SLP MMSE problem is formulated as
    \begin{align}
			&\min_{\substack{
      {\bf x}_{{\rm c}}[l],\,\tilde{\bf s}_{{\rm c}}[l],\\
      \boldsymbol{\Gamma}[l], \forall l \in\mathcal{L}
    }}  \sum_{l=1}^{L} {\mathbb{E}}_{{{\bf n}}}\left\{\left\|{{\boldsymbol{\Gamma}}}^{-1}[l] \left({{\bar{\bf H}}[l]{\bf x}_{{\rm c}}[l]+{\bar{\bf n}}[l]}\right)-{\tilde{\bf s}}_{{\rm c}}[l]\right\|^2_2\right\}\notag\\
			&\quad{\rm s.t.}\ \left\|{\bf x}_{{\rm c}}[l]\right\|^2_2 \leq P_{\rm T},\ \forall l \in\mathcal{L},\notag\\
			&\quad\quad\quad\tilde{ s}_{k}[l]\in\mathcal{D}_{k}[l],\ \forall k \in {\mathcal{ K}},\forall l \in\mathcal{L},\label{eq:CIMMSE-R problem}\\
			&\quad\quad\quad\gamma_{k}[l]>0,  \;\forall k \in {\mathcal{ K}},\forall l \in\mathcal{L},\notag
	\end{align}
    where ${{\boldsymbol{\Gamma}[l]}}=$${\rm diag}\{\gamma_{1}[l], ..., \gamma_{K}[l]\}$, $\bar{\bf H}[l]=$$[\bar{\bf h}_{1}[l],\ldots,\bar{\bf h}_{K}[l]]^T$, and $\bar{\bf n}[l]=$$[\bar{n}_{1}[l],\ldots,\bar{n}_{K}[l]]^T$. This problem can be decoupled into per-symbol optimization problems, which are then solved via an alternating optimization approach \cite{wang2024robust}, yielding a closed-form solution
    \begin{align}
		{{\bf x}}_{\rm{c}}^{\star}[l]=\eta^{\star}[l]{\bf P}[l]{\tilde{\bf s}}_{\rm{c}}[l],\ \
		\eta^{\star}[l] = \sqrt{{P_{\rm T}}/{\left\|{\bf P}[l]{\tilde{\bf s}}_{\rm{c}}[l]\right\|^2_2}},\label{eq:CIMMSE-R closedform}
    \end{align}
    with
    \begin{align}
		&{\bf P}[l] = \left({\bar{\bf H}}^H[l]{{\boldsymbol{\Psi}}}^2[l]{\bar{\bf H}}[l]+{\boldsymbol {\Phi}}[l]+\frac{{\sigma^2}\sum_{k=1}^{K}{\psi^2_{k}[l]}}{P_{\rm T}}{\bf I}_{N_{\rm T}}\right)^{-1}\notag\\
        &\quad\quad\quad\ {\bar{\bf H}}^H[l]{{\boldsymbol{\Psi}}}[l],\label{eq:CIMMSE Pmat}
		\\
		&{\boldsymbol {\Phi}}[l] = \sum^{K}_{k=1}{\psi^2_{k}[l]}{\beta}^2_{k}[l]{\bf E}_k,\label{eq:CIMMSE upsilon}\\
		&\tilde{\mathbf{s}}_{{\rm c}}[l]={\mathbf{s}}_{{\rm c}}[l]+\boldsymbol{\Lambda}_{\mu}[l]\boldsymbol{\delta}_{\mu}[l]+\boldsymbol{\Lambda}_{\nu}[l]\boldsymbol{\delta}_{\nu}[l],\label{eq:CIMMSE tildeS}
	\end{align}
    where $\boldsymbol{\Psi}[l]=\text{diag}\{\psi_{1}[l],\ldots,\psi_{K}[l]\}$, $\mathbf{E}_k=\mathbf{V}_k^T\mathbf{V}_k$, and $\frac{\psi_{k}[l]}{\eta[l]} = \frac{1}{\gamma_{k}[l]}$.
    When $\boldsymbol{\Psi}[l]$ is obtained, problem \eqref{eq:CIMMSE-R problem} can be reformulated as \cite{wang2024robust}
    \begin{align}
        \begin{aligned}
            &\min_{\boldsymbol{\delta}_{\mu}[l]\succeq{\bf 0},\boldsymbol{\delta}_{\nu}[l]\succeq{\bf 0}}\sum_{l=1}^{L}\tilde{\mathbf{s}}_{{\rm c}}^H[l]\boldsymbol{\Upsilon}[l]\tilde{\mathbf{s}}_{{\rm c}}[l]\\
            &{\rm s.t.} \ \tilde{\mathbf{s}}_{{\rm c}}[l]={\mathbf{s}}_{{\rm c}}[l]+\boldsymbol{\Lambda}_{\mu}[l]\boldsymbol{\delta}_{\mu}[l]+\boldsymbol{\Lambda}_{\nu}[l]\boldsymbol{\delta}_{\nu}[l],\forall l \in\mathcal{L},\\
            &\quad\ \  \boldsymbol{\Upsilon}[l]=({\bf I}_{K}-\boldsymbol{\Psi}[l]\bar{\bf H}[l]{\bf P}[l]),\forall l \in\mathcal{L}.\label{eq:rcimmse unif form}
        \end{aligned}
    \end{align}
    Given its strong resemblance to problems \eqref{eq:cizf-problem-upslion} and \eqref{eq:te design cimmse p1}, this problem can also be solved using the aforementioned SLPN.

    When the key variables $\psi_{k}[l]$, $\boldsymbol{\delta}_{\nu}[l]$, and $\boldsymbol{\delta}_{\mu}[l]$ are determined, the optimal transmit vector ${{\bf x}}_{\rm{c}}^{\star} [l]$ can be obtained in closed form via \eqref{eq:CIMMSE-R closedform}-\eqref{eq:CIMMSE tildeS}. Consequently, the subsequent challenge is to design a low-complexity neural network for the estimation of these variables.
    
    \subsection{Tensor Equivariance in Robust SLP Design}
    As formulated in \eqref{eq:CIMMSE-R problem}, the $\psi_{k}[l]$ is dependent on the inputs $\bar{\bf H}[l]$, ${\bf s}_{{\rm c}}[l]$, $\alpha_{k}[l]$, $\sigma^2$, and ${\bf m}_k$. Therefore, we can define a mapping to represent this relationship. For notational compactness, we first define 
    \begin{align}
        \begin{aligned}
            &{\bf A}\in\mathbb{R}^{K\times L},\ \text{where}\ [{\bf A}]_{k,l}=\alpha_{k}[l],\\
            &\boldsymbol{\Psi}=[\boldsymbol{\psi}[1],\ldots,\boldsymbol{\psi}[L]]\in\mathbb{R}^{K\times L},\\ &\boldsymbol{\psi}[l]=[\psi_{1}[l],\ldots,\psi_{K}[l]]^T\in\mathbb{R}^{K\times 1},\\
            &{\bf M}=[{\bf m}_1,\ldots,{\bf m}_K]^T\in\mathbb{R}^{K\times N_{\rm T}},\\ 
            &{\bf U}={\bf M}{\bf V}_{\rm T}^H\in\mathbb{C}^{K\times N_{\rm T}},\\
            &{\bf S}=[{\mathbf{s}}_{{\rm c}}[1],\ldots,{\mathbf{s}}_{{\rm c}}[L]]\in\mathbb{C}^{K\times L}.
        \end{aligned}\label{eq:rcimmse definition}
    \end{align}
    Then, we stack these components to obtain
    \begin{align}
    \begin{aligned}
        \tQ&=[{\bf H},{\bf U}]_3\in\mathbb{C}^{K\times N_{\rm T}\times 2}, \\ \tG&=[\real\{{\bf S}\},\imaginary\{{\bf S}\},{\bf A}]_3 \in\mathbb{R}^{K\times L\times 3}.
    \end{aligned}
        \label{eq:rcimmse get E G}
    \end{align}
    Finally, we define the mapping as
    \begin{align}
        G_{\rm P1}(\tQ,\tG,\sigma^2)=\boldsymbol{\Psi}^\star.\label{eq:rcimsme psi mapping}
    \end{align}
    This mapping satisfies the following property
    \begin{ppn}\label{ppn: te for cimmser problem}
    When \eqref{eq:CIMMSE-R problem} has a unique optimal solution for $\boldsymbol{\Psi}$, i.e., $G_{\rm P1}(\tQ,\tG,\sigma^2)=\boldsymbol{\Psi}^\star$, for any $\pi_K\in\mathbb{S}_K$, $\pi_{N_{\rm T}}\in\mathbb{S}_{N_{\rm T}}$, and $\pi_L\in\mathbb{S}_L$, the following equations hold
    \begin{subequations}
        \begin{align}
        &G_{\rm P1}(\pi_{K}\circ_1\tQ,\pi_{K}\circ_1\tG,\sigma^2)=\pi_{K}\circ_1\boldsymbol{\Psi}^\star,\label{eq:cimmser TE a}\\ 
        &G_{\rm P1}(\pi_{N_{\rm T}}\circ_2\tQ,\tG,\sigma^2)=\boldsymbol{\Psi}^\star,\label{eq:cimmser TE b}\\
        &G_{\rm P1}(\tQ,\pi_{L}\circ_2\tG,\sigma^2)=\pi_{L}\circ_2\boldsymbol{\Psi}^\star.\label{eq:cimmser TE c}
    \end{align}
    \end{subequations}
    \end{ppn}
    \begin{pf}
	The proof follows a similar procedure to that in \ppnref{ppn: te for cizf problem}.
    \end{pf}

    As for the mapping to $\boldsymbol{\delta}_{\nu} [l]$ and $\boldsymbol{\delta}_{\mu}[l]$, since \eqref{eq:rcimmse unif form} and \eqref{eq:cizf-problem-upslion} share nearly identical forms, we can first employ the same method as in \ref{sec:gradient mapping} to derive its KKT conditions, and we define
    \begin{align}
        {\bf B}_{\rm c}[l]&=\big[\boldsymbol{\Lambda}_{\mu}^H[l]\boldsymbol{\Upsilon}[l]{\mathbf{s}}_{{\rm c}}[l],\boldsymbol{\Lambda}_{\nu}^H[l]\boldsymbol{\Upsilon}[l]{\mathbf{s}}_{{\rm c}}[l]\big]\in\mathbb{C}^{K\times2},\label{eq:cal gradient robust 1}\\
        {\tC}_{\rm c}[l]&=\big[\boldsymbol{\Lambda}_{\mu}^H[l]\boldsymbol{\Upsilon}[l]\boldsymbol{\Lambda}_{\mu}[l],\boldsymbol{\Lambda}_{\mu}^H[l]\boldsymbol{\Upsilon}[l]\boldsymbol{\Lambda}_{\nu}[l],\notag\\&\boldsymbol{\Lambda}_{\nu}^H[l]\boldsymbol{\Upsilon}[l]\boldsymbol{\Lambda}_{\mu}[l],\boldsymbol{\Lambda}_{\nu}^H[l]\boldsymbol{\Upsilon}[l]\boldsymbol{\Lambda}_{\nu}[l]\big]_3\in\mathbb{C}^{K\times K\times 4}.\label{eq:cal gradient robust 2}
    \end{align}
    Then ${\tB}_{\rm c}$ and ${\tC}_{\rm c}$ are definde according to \eqref{eq:cal gradient 3} and \eqref{eq:cal gradient 4}, and $\tD$ is defined as \eqref{eq:definition of D}. Finally, we define the mapping as follows
    \begin{align}
        G_{\rm P2}(\tB_{\rm c},\tC_{\rm c})=\tD^\star,\label{eq:rcimmse delta mapping}
    \end{align}
    which is exactly the same as \eqref{eq:the mapping} and also satisfies the TE described in \eqref{eq:cizf pos TE a} and \eqref{eq:cizf pos TE b}.

	\subsection{{TE-Based Low-Complexity Robust SLP Framework}}\label{sec:CIMMSERnet}
    Leveraging the TE analyzed in the previous subsection, we then propose a  DL-based low-complexity framework for robust SLP, with the following two networks to determine the auxiliary variables $\boldsymbol{\Psi}$ and the perturbation factors $\tD$:

    \subsubsection{Network for Auxiliary Variables}
    As shown in Fig. \ref{pic:TECISBR}, we design the network RSLPN-A to approximate the mapping $G_{\rm P1}$, ensuring that the network is structured to fully exploit the TE of $G_{\rm P1}$. The input is constructed as follows
     \begin{align}
         \begin{aligned}
         &{\tX}=[\real(\tilde{\tQ}),\imaginary(\tilde{\tQ}),\tilde{\tG},\sigma^2{\bf 1}]_4\in\mathbb{R}^{K\times N_{\rm T}\times L\times D_I},\label{eq:rcimmse get X 2}\\
             &\tilde{\tG}=\{{\tG}\}_{2,N_{\rm T}}\in\mathbb{R}^{K\times N_{\rm T}\times L\times 3}, \\ &\tilde{\tQ}=\{{\tQ}\}_{3,L}\in\mathbb{C}^{K\times N_{\rm T}\times L\times 2},
         \end{aligned}
     \end{align}
     where $D_I=8$, and $\{\cdot\}_{p,q}$ denotes the operation that replicates the input tensor $q$ times along a newly inserted $p$-th dimension. 
    \begin{figure}[t]
	\centering
	\includegraphics[width=1.8 in]{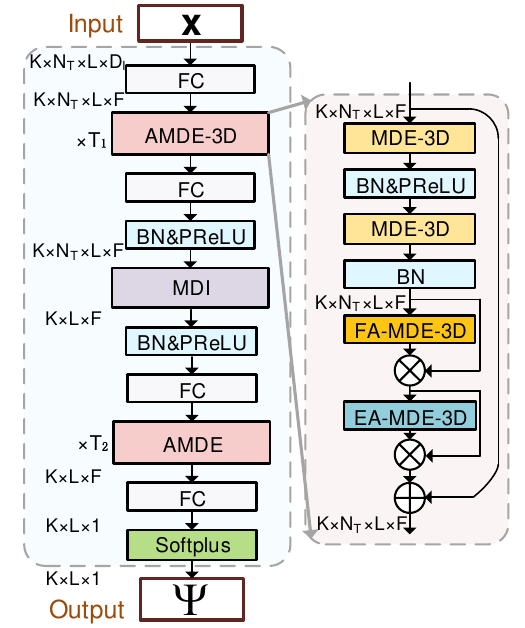}
	\caption{The architecture of RSLPN-A.}
	\label{pic:TECISBR}
    \end{figure}
    To process the 4-D input tensor with TE across its first three dimensions, the network employs a specialized 3-D AMDE module, denoted as AMDE-3D. Within this module, the MDE and RMDE layers are designed to enforce equivariance along the first three dimensions, while the FA-MDE component is configured with $\mathcal{D}=\{1,2\}$. The transmit antenna dimension is processed by the MDI layer in \cite{11049893} to capture its invariance. In addition, A softplus layer is used as the output layer to ensure that $\boldsymbol{\Psi}$ satisfies the positivity constraint.

    \subsubsection{Network for Perturbation Factors}
    Since the mapping $G_{\rm P2}$ has the same form as the mapping $G$ in \eqref{eq:the mapping} and satisfies the same TE, it can be approximated using the SLPN proposed in Section \ref{LC robust SINR}. For distinction, we refer to this network as RSLPN-B in this section.

    \subsubsection{Overall Structure of the RSLP Framework}
    The proposed framework operates in two sequential stages: first, RSLPN-A provides an estimate of the matrix $\boldsymbol{\Psi}$; second, this estimate is used to construct the input for RSLPN-B, which subsequently yields the final output $\tD$.
    In detail, $\tQ$ and $\tG$ are computed from \eqref{eq:rcimmse definition}-\eqref{eq:rcimmse get E G} and employed to derive $\tX$ using \eqref{eq:rcimmse get X 2}. This $\tX$ serves as the input to the RSLPN-A network, which generates the estimate $\boldsymbol{\Psi}$. Following this, the matrix ${\bf P}[l]$ is obtained using \eqref{eq:CIMMSE Pmat}-\eqref{eq:CIMMSE upsilon}. Then, $\boldsymbol{\Upsilon}[l]$ is computed via \eqref{eq:rcimmse unif form}, which is then normalized by its Frobenius norm. The terms ${\tB}_{\rm c}$ and ${\tC}_{\rm c}$ are then obtained from \eqref{eq:cal gradient robust 1}, \eqref{eq:cal gradient robust 2}, \eqref{eq:cal gradient 3}, \eqref{eq:cal gradient 4}, and $\tB$ and $\tC$ are defined in \eqref{eq:cal input 1}-\eqref{eq:cal input 2}. Then $\tD$ is determined by RSLPN-B, and the variables $\hat{\boldsymbol{\delta}}_{\mu}[l]$ and $\hat{\boldsymbol{\delta}}_{\nu}[l]$ are determined from \eqref{eq:relu D}-\eqref{eq:hatD to delta}, while $\tilde{{\bf s}}_{\rm c}[l]$ is calculated via \eqref{eq:CIMMSE tildeS}. Finally, the transmit vector ${\bf x}_{\rm c}[l]$ is given by \eqref{eq:CIMMSE-R closedform}. The MSE loss function, as defined in \eqref{eq:loss mse}, is utilized for training both subnetworks.

	\section{Numerical Results}\label{Section 6}
    In this section, we evaluate the performance of the proposed methods via Monte Carlo simulations.
    To generate channels compliant with 3GPP standards, we utilize the widely adopted QuaDRiGa channel simulator \cite{6758357}.
    We consider a MIMO system where the BS is equipped with an $N_{\rm T}\times 1$ uniform linear array (ULA) of type `3gpp-3d', serving $K$ single-antenna users, each equipped with an `omni' antenna. Users are randomly located within a 500-meter radius in a 120-degree sector facing the BS. The carrier frequency is 3.5 GHz, and the propagation environment corresponds to the `3GPP\_38.901\_UMa\_NLOS' scenario \cite{3gpp_tr_38901_v190}. Shadow fading and path loss are not considered. The channel is normalized such that $\mathbb{E}\{{\rm tr}({\bf H}{\bf H}^H)\} = KN_{\rm T}$. The SNR is defined as ${\rm SNR} = P_{\rm T}/\sigma^2$. The imperfect channels were generated according to the widely adopted \textit{a posteriori} model in \eqref{posteriori}, where $\{{\bf m}_k\}^K_{k=1}$ was estimated from the channel realizations generated by the QuaDRiGa channel simulator \cite{8694866}. For simplicity, we assume $\alpha_{k,l} = \alpha,\forall k\in\mathcal{K},\forall l\in\mathcal{L}$, i.e., the same $\alpha$ is shared across all users and symbols within a transmission block. Under the predefined channel scenario configuration, each channel realization in the dataset is independently generated by QuaDRiGa, with random user distributions, scatterer distributions, and channel statistics to ensure diversity across channel realizations.

    {This section compares the following schemes:}
    \begin{itemize}
		\item `\textbf{ZF}' and `\textbf{MMSE}' \cite{Li2021,1391204}: The ZF and MMSE precoding schemes with symbol-level power constraints.
		\item `\textbf{CIZF}' and `\textbf{CIMMSE}'\cite{9910472,8815429}: The conventional SLP solutions to problems \eqref{eq:cizf-problem-persymbol} and \eqref{eq:cimmse-problem}, respectively.
        \item `\textbf{CIZF-CF}' \cite{8465957}: The approximate closed-form solution for CIZF problem.
        \item `\textbf{CIZF-DL}' and `\textbf{CIMMSE-DL}': The TENN-based SLP framework proposed in Section \ref{perfect csi slp}. The network is configured with $T=4$, $F=4$.
        \item  `\textbf{RCIMMSE}': The solutions to the robust MMSE SLP problems \eqref{eq:CIMMSE-R problem} proposed in \cite{wang2024robust}. 
        \item `\textbf{RCIMMSE-DL}': The TENN-based SLP neural networks proposed in Section \ref{sec:CIMMSERnet}. The RSLPN-A is configured with $T_1=2, T_2=2$, $F=16$, and the RSLPN-B is configured with $T=2$, $F=16$.
	\end{itemize}

    \subsection{Training Details}
    For the SLPN in Section \ref{perfect csi slp}, the dataset consisted of 100,000 independent channel realizations, with 90,000 used for training and 10,000 for testing, each associated with a transmit symbol sequence of length $L=100$. The corresponding perturbation factors under the CIZF and CIMMSE criteria were then computed as labels. To improve robustness across SNRs, the CIMMSE labels were generated over a set of SNR values $\{0, 5, 10, 15, 20, 25, 30\}$ dB. Separate networks were trained for CIZF and CIMMSE for 800 and 400 epochs, respectively.
    For the RSLPN in Section \ref{sec:CIMMSERnet}, a dataset of 60,000 distinct channel realizations (55,000 for training and 5,000 for testing) was generated, each associated with the corresponding ${\bf M}$ and 50 different transmit symbols. The corresponding auxiliary variables and perturbation factors were used as labels, generated over SNR values $\{10, 15, 20, 25, 30, 35, 40\}$ dB. The training was performed sequentially: RSLPN-A was first trained to estimate $\boldsymbol{\Psi}$, which is then used to train RSLPN-B. The two subnets of RSLPN were each trained for 300 epochs.
    
    All networks were trained using the Adam optimizer with a batch size of 400. A two-stage learning rate schedule was adopted: $5 \times 10^{-3}$ for the first half of the training epochs and $5 \times 10^{-4}$ for the second half.
    Although the networks were trained on datasets with a fixed number of users $K$ and symbol block length $L$, their design inherently provides generalization capability to other configurations. Moreover, the test set consists of channel realizations that were not encountered during training, characterized by different channel environments, user locations, and distinct transmit symbols.

    \subsection{{Complexity Comparison}}\label{sec:complexity comparison}

\begin{table}[!t]
    \centering
    \caption{Per Symbol Computational Complexity }
     \label{tab:precoding-complexity}
    \resizebox{0.7\columnwidth}{!}{%
        \begin{tabular}{lcc}
        \toprule
        \textbf{Methods} & \textbf{Complexity Order} \\
        \midrule
         ZF  & $\mathcal{O}(KN_{\rm T} )$  \\
         MMSE & $\mathcal{O}(KN_{\rm T})$   \\
         CIZF & $\mathcal{O}\bigl(KN_{\rm T}+KN_{\rm T}N_L+N_{\rm T}N_L^3\bigr)$ \\
         CIZF-CF  & $\mathcal{O}\bigl(KN_{\rm T}+M_{\rm CF}^3\bigr)$  \\
         CIMMSE & $\mathcal{O}\bigl(KN_{\rm T}+K^2N_L+KN_L^3\bigr)$ \\
         CIZF-DL & $\mathcal{O}(KN_{\rm T}+KF^2\bigr)$ \\
        CIMMSE-DL & $\mathcal{O}\bigl(KN_{\rm T}+KF^2\bigr)$ \\ 
        RCIMMSE&$\mathcal{O}\bigl((N_{\rm T}^3+KN_L^3)I\bigr)$\\
        RCIMMSE-DL&$\mathcal{O}\bigl(N_{\rm T}^3+KN_{\rm T}F^2\bigr)$\\
         \bottomrule
        \end{tabular}%
    }
    \end{table}
    
    Table \ref{tab:precoding-complexity} presents the dominant order complexity for various precoding algorithms. Let $K$, $N_{\rm T}$, and $N_L$ denote the number of UEs, BS antennas, and main loop iterations for the active set algorithm, respectively. $M_{\rm CF}$ is the estimated size of the inactive constraint set for CIZF-CF, $F$ represents the number of hidden layer neurons, and $I$ is the number of RCIMMSE iterations (which stops when the MSE difference between two consecutive iterations falls below $10^{-4}$ or after 400 iterations). We assume $L\gg N_{\rm T}\ge K$, and $2K>N_L,M_{\rm CF}$. In the non-robust scenario, $K>F$, while in the robust scenario, $N_{\rm T}\approx F$. 
    The proposed CIZF-DL and CIMMSE-DL frameworks employ the low-complexity SLPN, thereby avoiding the iterations required for solving NNLS problems. As a result, they overcome the high-order $N_{\rm T} N_{L}^{3}$ complexity and achieve online computational efficiency that scales linearly with both $K$ and $N_{\rm T}$, consistent with LP. The complexity of RCIMMSE is primarily determined by the number of iterations $I$, whereas the proposed RCIMMSE-DL eliminates this and significantly reduces computational load.
    Furthermore, the proposed method generalizes effectively across both $K$ and $L$, allowing a single trained network to adapt to different $K$ and $L$ configurations, thereby avoiding the overhead of training and storing multiple networks.

    \subsection{Performance of SLP with Perfect CSI}

            \begin{figure*}[!t]
		\centering
		\subfigure[$N_{\rm T}=12$, $K=12$, 4-QAM]{
			\begin{minipage}[t]{0.31\linewidth}
				\centering
				\includegraphics[width=1\textwidth]{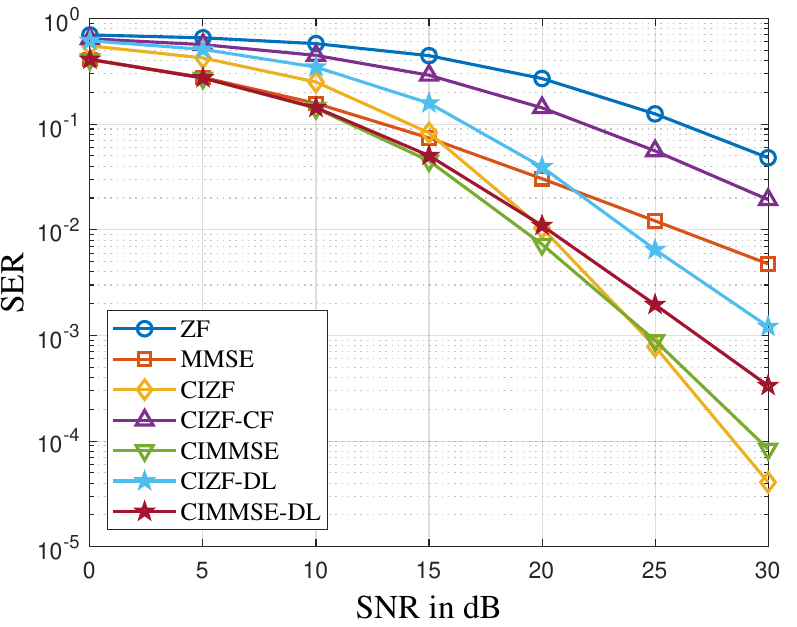}
			\end{minipage}%
		}%
		\subfigure[$N_{\rm T}=14$, $K=12$, 4-QAM]{
			\begin{minipage}[t]{0.31\linewidth}
				\centering
				\includegraphics[width=1\textwidth]{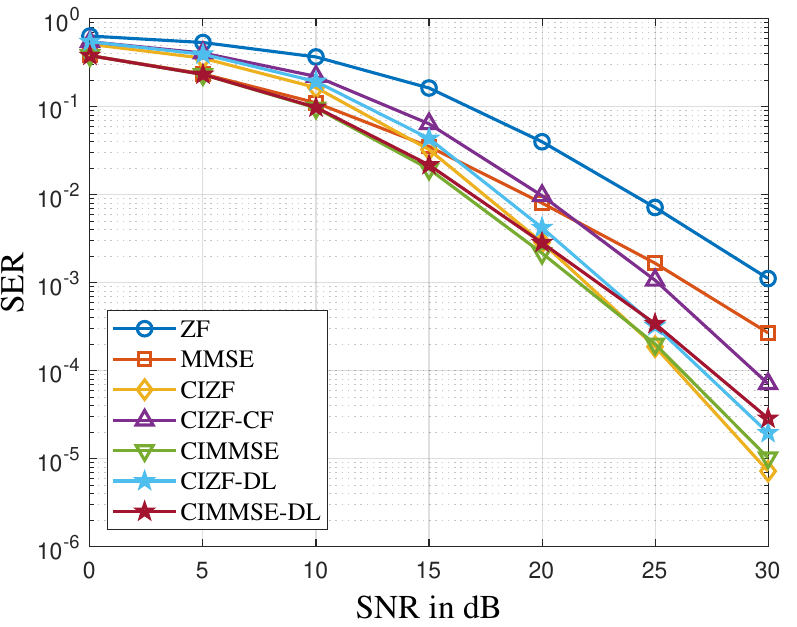}
			\end{minipage}
            % \label{pic:Perfect CSI L=1 b}
		}
            \subfigure[$N_{\rm T}=14$, $K=12$, 16-QAM]{
			\begin{minipage}[t]{0.31\linewidth}
				\centering
				\includegraphics[width=1\textwidth]{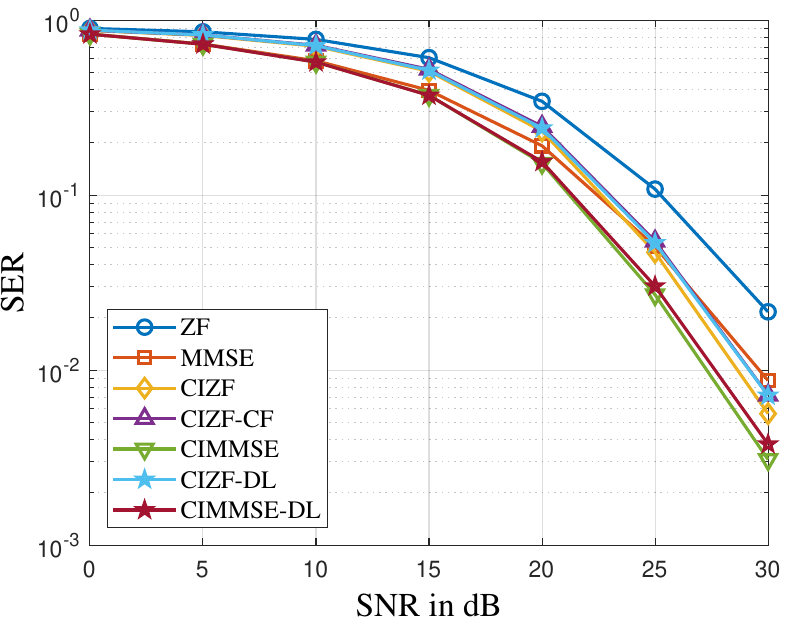}
			\end{minipage}
            % \label{pic:Perfect CSI L=1 b}
		}
		\centering
        % \vspace*{-1mm}
		\caption{The SER versus SNR under different antenna configurations and modulation schemes, $L=100$.}
		\label{pic:Perfect CSI L=1}
	\end{figure*}

    Fig. \ref{pic:Perfect CSI L=1} compares the SER performance of various schemes. As can be observed, both CIZF-DL and CIMMSE-DL consistently outperform conventional ZF and MMSE, while also achieving gains over CIZF-CF. Specifically, CIMMSE-DL achieves a lower SER than CIZF in the low-SNR regime, while CIZF-DL surpasses MMSE at high SNRs. For instance, with $N_{\rm T} = K = 12$ and 4-QAM, CIMMSE-DL requires approximately 6 dB less SNR than MMSE to achieve an SER of $10^{-2}$. Under the same conditions, CIZF-DL delivers about 3 dB of SNR gain over MMSE and more than 6 dB over CIZF-CF. These results affirm that the proposed SLP framework effectively retains the performance benefits of SLP.

    \begin{figure*}[!t]
    \centering
    \begin{minipage}[b]{0.30\linewidth}
        \centering
        \includegraphics[width=\textwidth]{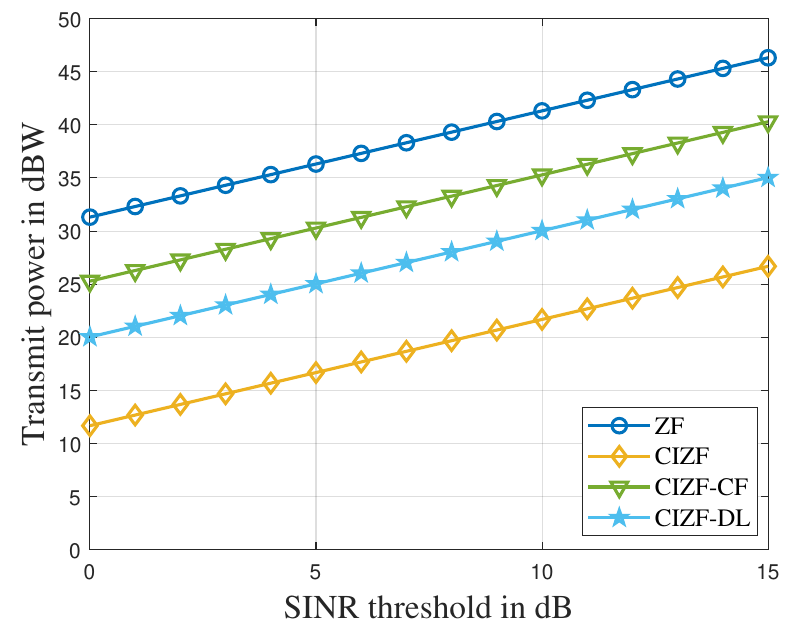}
        % \captionsetup{width=0.3\linewidth}
        \vspace*{-7mm}
        \caption{Total transmit power versus SINR threshold, $N_{\rm T}=12$, $K=12$, 4-QAM.}
        \label{TransmitPower_SINRthreshold}
    \end{minipage}%
    \hspace{0.02\linewidth}
    \begin{minipage}[b]{0.30\linewidth}
        \centering
        \includegraphics[width=\textwidth]{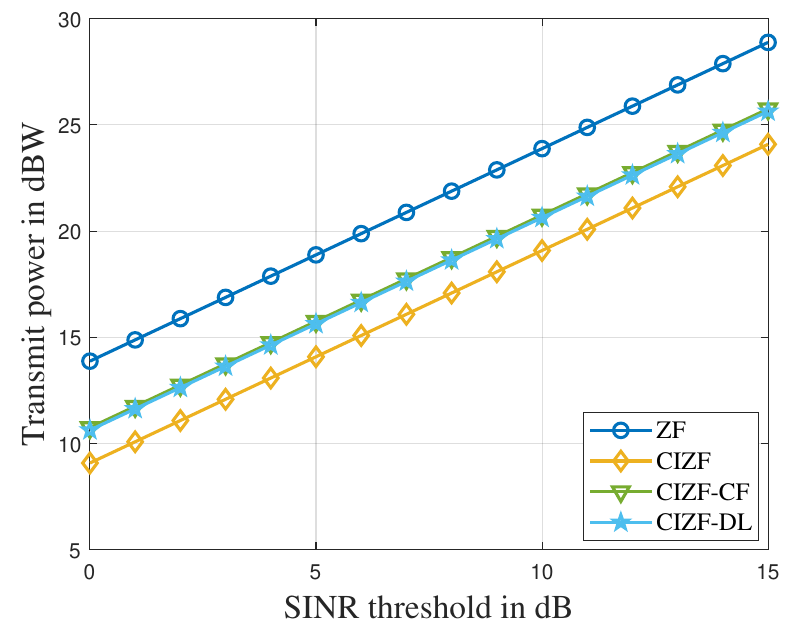}
        \vspace*{-7mm}
        \caption{Total transmit power versus SINR threshold, $N_{\rm T}=14$, $K=12$, 4-QAM.}
        \label{TransmitPower_SINRthreshold_new}
    \end{minipage}%
    \hspace{0.02\linewidth}
    \begin{minipage}[b]{0.30\linewidth}
        \centering
        \includegraphics[width=\textwidth]{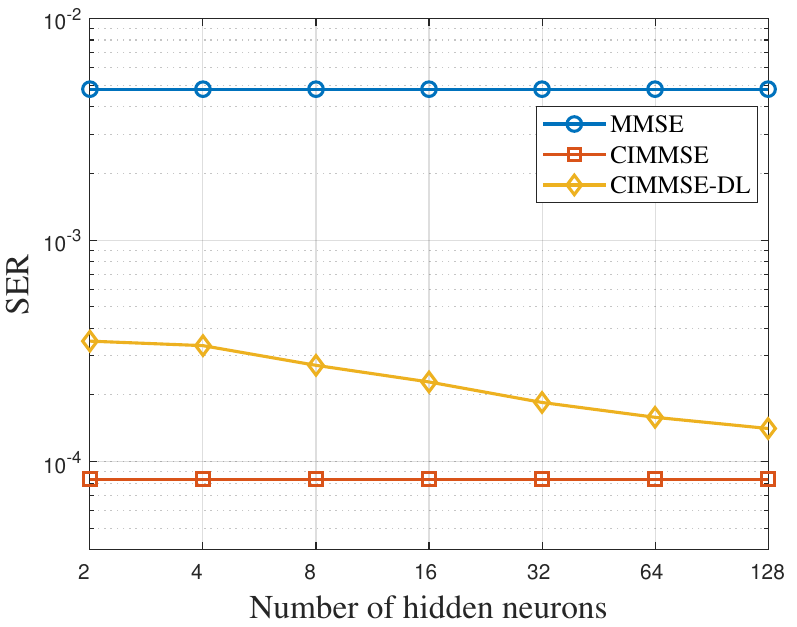}
        \vspace*{-7mm}
        \caption{SER across network configurations, $N_{\rm T}=12$, $K=12$, 4-QAM, 30dB, $T=4$.}
        \label{SERvsD2}
    \end{minipage}
\end{figure*}

                \begin{table}[t]
    \centering
    \caption{Average Execution Time per Symbol (s)}
     \label{tab:precoding-time-robust}
    \resizebox{0.7\columnwidth}{!}{%
        \begin{tabular}{lccc}
        \toprule
        \textbf{Methods} & \textbf{CPU} & \textbf{GPU} \\
        \midrule
         RCIMMSE      & $1.7\times 10^{-1}$ & $3.5\times 10^{-1}$  \\
        RCIMMSE-DL               & $3.1\times 10^{-4}$ & $2.3\times 10^{-4}$  \\ 
         \bottomrule
        \end{tabular}%
    }
    \end{table}

    Figs. \ref{TransmitPower_SINRthreshold} and \ref{TransmitPower_SINRthreshold_new} depict the total transmit power as a function of the SINR threshold, defined as $\gamma^2/\sigma^2$. Under a given SINR threshold, a reduction in transmit power corresponds to improved energy efficiency. As evidenced, the proposed CIZF-DL scheme achieves substantial power gains of approximately 11.3 dB over ZF precoding and 5.3 dB over CIZF-CF in the $N_{\rm T}=12$, $K=12$ scenario. Furthermore, it delivers transmit power comparable to the optimal CIZF solution in the $N_{\rm T}=14$, $K=12$ scenario. These results clearly demonstrate the superior power efficiency of our CIZF-DL approach, which is a critical advantage for practical implementations.

    \begin{figure*}[!t]
    \centering
    \begin{minipage}[b]{0.30\linewidth}
        \centering
        
        \includegraphics[width=\textwidth]{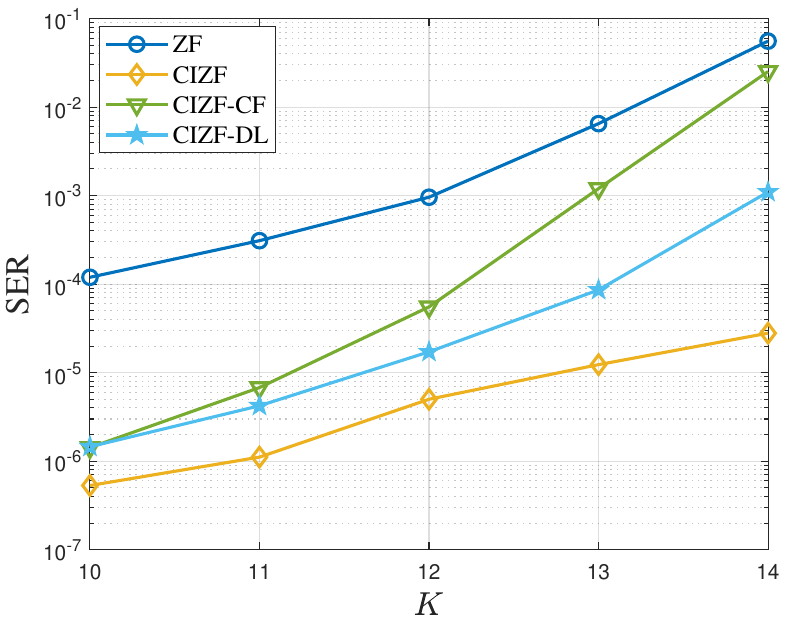}
        % \captionsetup{width=0.3\linewidth}
        \vspace*{-7mm}
        \caption{Generalization performance with respect to $K$ for the network trained under scenario $N_{\rm T}{=}14$, $K{=}12$, and $L{=}100$.}
        \label{gen-figure1}
    \end{minipage}%
    \hspace{0.02\linewidth}
    \begin{minipage}[b]{0.30\linewidth}
        \centering
        \includegraphics[width=\textwidth]{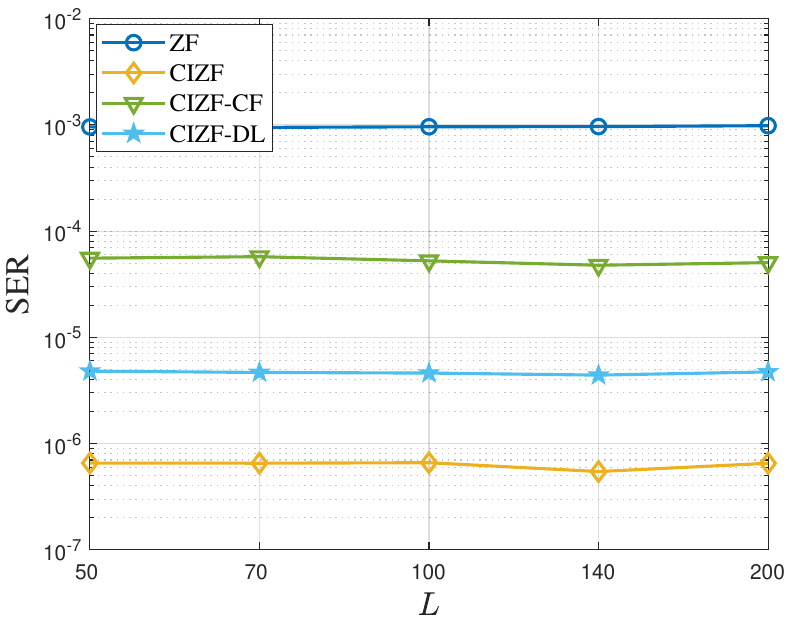}
        \vspace*{-7mm}
        \caption{Generalization performance with respect to $L$ for the network trained under scenario $N_{\rm T}{=}14$, $K{=}12$, and $L{=}100$.}
        \label{gen-figure2}
    \end{minipage}%
    \hspace{0.02\linewidth}
    \begin{minipage}[b]{0.30\linewidth}
        \centering
        \includegraphics[width=\textwidth]{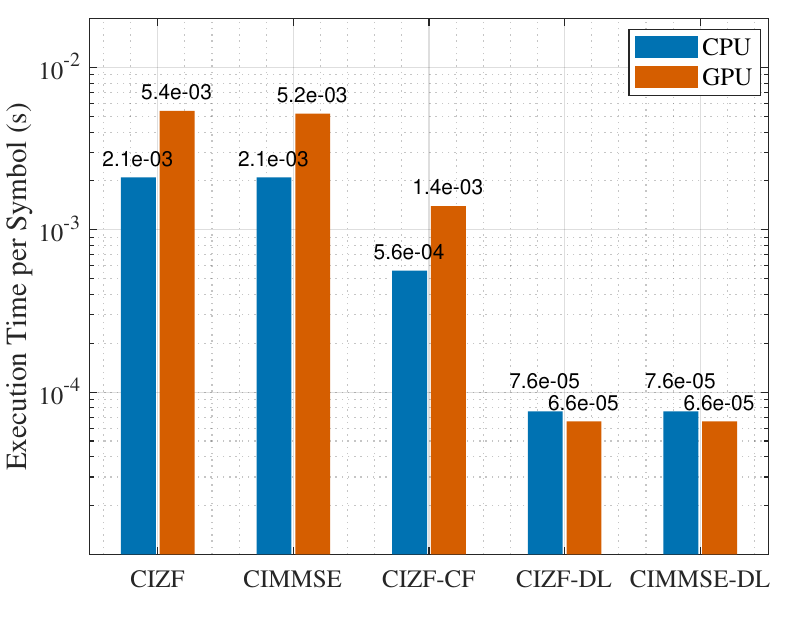}
        \vspace*{-7mm}
        \caption{Average execution time per symbol, $N_{\rm T}=14$, $K=12$, $L=100$, $\text{SNR}=30$dB, 4-QAM.}
        \label{runtime_perfectCSI}
    \end{minipage}
\end{figure*}

    The SER of the proposed network generally decreases as the size of the hidden layer increases, converging towards the optimal achievable performance. Fig. \ref{SERvsD2} depicts the SER of CIMMSE-DL as a function of its hidden layer size $F$. With increasing $F$, the SER steadily decreases, approaching the performance of CIMMSE and eventually reaching a closely comparable level. This trend demonstrates the capability of our proposed architecture to effectively approximate sophisticated SLP solutions. \figref{gen-figure1} and \figref{gen-figure2} demonstrate the generalization capability of the proposed approach. We train our network in scenario $N_{\rm T}=14$, $K=12$, $L=100$, and directly apply it to various scenarios. As can be observed, the proposed method exhibits consistently outstanding performance, highlighting its robust practical utility.
    
    \figref{runtime_perfectCSI} shows the average execution times implemented in PyTorch on Intel(R) Xeon(R) Platinum 8336C CPU and NVIDIA GeForce RTX 4090 GPU.
    The proposed CIZF-DL and CIMMSE-DL methods exhibit significantly lower execution times on both platforms compared to conventional SLP precodings. Specifically, CIZF-DL consumes merely 3.62\% of the CPU runtime and 1.22\% of the GPU runtime required by CIZF. Similarly, the CIMMSE-DL demands only approximately 3.62\% of the CPU and 1.27\% of the GPU execution time relative to CIMMSE. This substantial speed-up is attributed to the ability of the proposed methods to process transmit symbols in blocks, which greatly enhances operational efficiency.

                \begin{figure}[t]
		\centering
		\includegraphics[width=0.7\columnwidth]{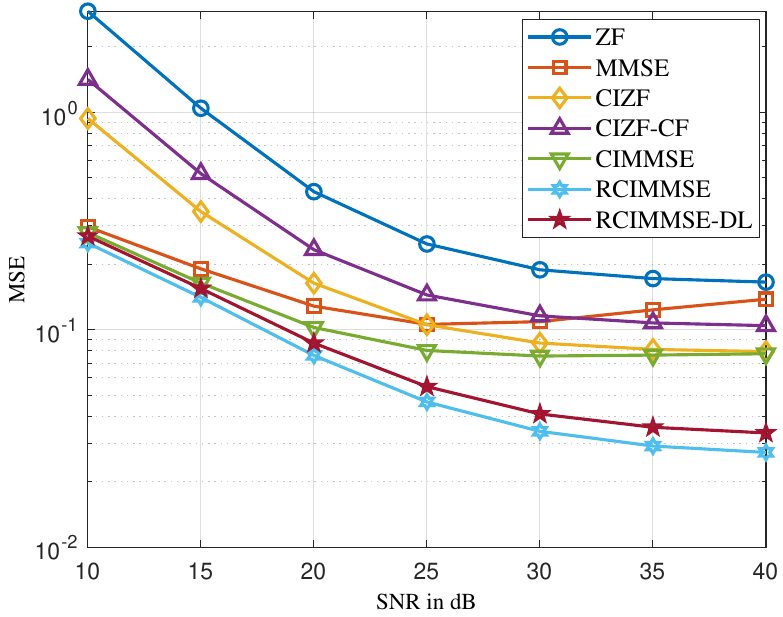}
        \vspace*{-3mm}
		\caption{MSE vs SNR, $N_{\rm T}=14$, $K=12$, $\alpha=0.995$, QPSK.}
		\label{ULA MSE}	
	\end{figure}
	
\subsection{Performance of Robust SLP}

 Fig. \ref{ULA MSE} and \ref{ULA SER} compare the MSE and SER performance, respectively, for QPSK downlink transmission with $\alpha=0.995$, $N_{\rm T}=14$, and $K=12$. Fig. \ref{ULA MSE} shows that the RCIMMSE-DL achieves near-optimal MSE performance, consistently outperforming the other compared methods across the evaluated SNR range.  Fig. \ref{ULA SER} indicates that the SER performance of RCIMMSE-DL nearly matches that of RCIMMSE. Specifically, at an SNR of 40 dB, the SER for the CIMMSE precoding is approximately $4.7\times 10^{-3}$, while RCIMMSE-DL achieves a significantly lower SER of $3.2\times 10^{-4}$, very close to the RCIMMSE result of $2.4\times 10^{-4}$. This substantial improvement becomes increasingly pronounced at higher SNRs.

	\begin{figure}[!t]
		\centering
		\includegraphics[width=0.7\columnwidth]{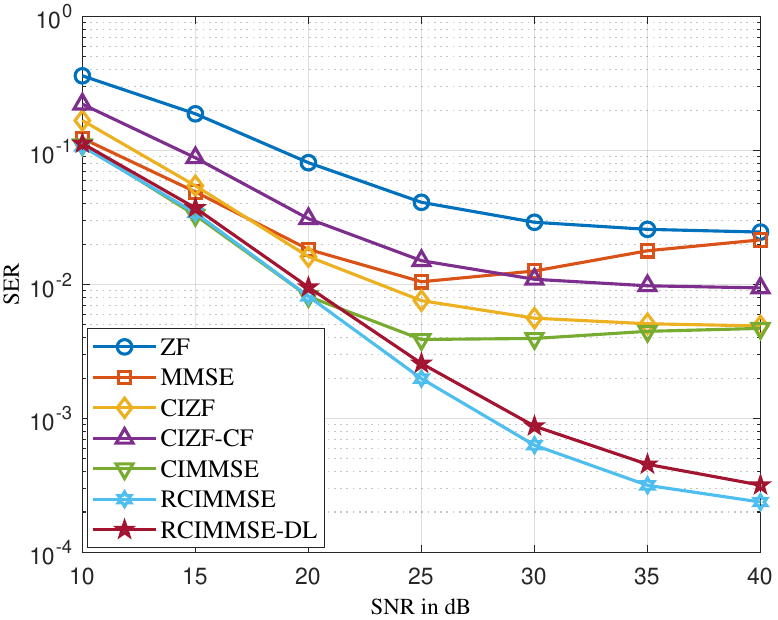}
        \vspace*{-3mm}
		\caption{SER vs SNR, $N_{\rm T}=14$, $K=12$, $\alpha=0.995$, QPSK.}
		\label{ULA SER}
	\end{figure}

    Table \ref{tab:precoding-time-robust} compares the average execution times of the RCIMMSE and RCIMMSE-DL methods. The simulations were configured with $N_{\rm T}=14$, $K=12$, $L=50$, $\alpha=0.995$, $\text{SNR}=30$dB, and QPSK modulation. A significant advantage in execution time for the proposed RCIMMSE-DL is clearly observable. Specifically, compared to the RCIMMSE method, the execution times of RCIMMSE-DL on the CPU and GPU are merely 0.18\% and 0.07\%, respectively. This demonstrates that our proposed method drastically reduces online computation time while maintaining excellent performance.

	\section{Conclusion}\label{Section 7}
        This paper proposed a framework based on TENN to address the high computational complexity of SLP problems. Leveraging the NNLS forms of the SLP problems and their closed-form solutions, we analyzed the optimality condition to establish a mapping from the available information to the perturbation factors, and then revealed the inherent TE within this mapping. Utilizing this TE, we proposed an attention-based tensor-equivariant neural network, based on which a low-complexity SLP framework is developed. Subsequently, considering imperfect CSI scenarios, we extended the proposed framework to the robust MMSE problem. Simulation results demonstrate the effectiveness of the proposed frameworks in terms of transmit power, SER,  execution time, and online computational complexity. Future work includes the joint design of SLP and receiver demodulation, aiming for further enhancements in system performance.

	\appendices
    \section{Proof of Proposition \ref{ppn: te for cizf problem}} \label{proof: te for cizf problem} 
    Define $M_l(\tB_{\rm c},\tC_{\rm c},\tD)=\nabla_{\boldsymbol{\delta}_{\mu}[l]}\mathcal{L}({\boldsymbol{\delta}}_{\mu}[l],{\boldsymbol{\delta}}_{\nu}[l],{\boldsymbol{\lambda}}_{\mu}[l],{\boldsymbol{\lambda}}_{\nu}[l])$, we first prove 
    \begin{align}
        \begin{aligned}
        &M_l(\pi_K\circ_1\tB_{\rm c},\pi_K\circ_{[1,2]}\tC_{\rm c},\pi_K\circ_1\tD)\\&=\pi_K\circ_1M_l(\tB_{\rm c},\tC_{\rm c},\tD).\label{eq:proof A 1}
    \end{aligned}
    \end{align}
    We define the permutation matrix $\boldsymbol{\Pi}$ to represent the permutation $\pi$ in the first dimension of a tensor. Each row and column of the matrix $\boldsymbol{\Pi}$ contains a single 1, with all other entries being 0, and $\boldsymbol{\Pi\Pi}^T={\bf I}$. Then, for $\pi_K\circ_1\tD,\pi_K\circ_{[1,2]}\tC_{\rm c},\pi_K\circ_1\tB_{\rm c}$, the variables during the $l$-th symbols can be expressed as ${\mathbf{s}}_{{\rm c}}'[l]={\boldsymbol{\Pi}\mathbf{s}}_{{\rm c}}[l],$ $\boldsymbol{\delta}_{\mu}'[l]=\boldsymbol{\Pi}\boldsymbol{\delta}_{\mu}[l],$ $\boldsymbol{\delta}_{\nu}'[l]=\boldsymbol{\Pi}\boldsymbol{\delta}_{\nu}[l],$ $\boldsymbol{\Lambda}_{\mu}'[l]=\boldsymbol{\Pi}\boldsymbol{\Lambda}_{\mu}[l]\boldsymbol{\Pi}^T,$ $\boldsymbol{\Lambda}_{\nu}'[l]=\boldsymbol{\Pi}\boldsymbol{\Lambda}_{\nu}[l]\boldsymbol{\Pi}^T$, ${\mathbf{H}}'={\boldsymbol{\Pi}\mathbf{H}}$. On this basis, we have $\boldsymbol{\Upsilon}'={\boldsymbol{\Pi}}\boldsymbol{\Upsilon}{\boldsymbol{\Pi}}^T$, $\tilde{\mathbf{s}}_{{\rm c}}'[l]={\boldsymbol{\Pi}}\tilde{\mathbf{s}}_{{\rm c}}[l]$. For simplicity, we temporarily drop $l$, and thus we have
    \begin{align}
            &M_l(\pi_K\circ_1\tB_{\rm c},\pi_K\circ_{[1,2]}\tC_{\rm c},\pi_K\circ_1\tD)\notag\\
            &=2\boldsymbol{\Pi}\boldsymbol{\Lambda}_{\mu}^H\boldsymbol{\Pi}^T{\boldsymbol{\Pi}}\boldsymbol{\Upsilon}{\boldsymbol{\Pi}}^T{\boldsymbol{\Pi}}{\mathbf{s}}_{{\rm c}}\!+\!2\boldsymbol{\Pi}\boldsymbol{\Lambda}_{\mu}^H\boldsymbol{\Pi}^T\boldsymbol{\Pi}\boldsymbol{\Upsilon}\boldsymbol{\Pi}^T\boldsymbol{\Pi}\boldsymbol{\Lambda}_{\mu}\boldsymbol{\Pi}^T\boldsymbol{\Pi}\boldsymbol{\delta}_{\mu}\notag\\
            &\qquad+2\boldsymbol{\Pi}\boldsymbol{\Lambda}_{\mu}^H\boldsymbol{\Pi}^T\boldsymbol{\Pi}\boldsymbol{\Upsilon}\boldsymbol{\Pi}^T\boldsymbol{\Pi}\boldsymbol{\Lambda}_{\nu}\boldsymbol{\Pi}^T\boldsymbol{\Pi}\boldsymbol{\delta}_{\nu}\\&=2\boldsymbol{\Pi}\boldsymbol{\Lambda}_{\mu}^H\boldsymbol{\Upsilon}{\mathbf{s}}_{{\rm c}}+2\boldsymbol{\Pi}\boldsymbol{\Lambda}_{\mu}^H\boldsymbol{\Upsilon}\boldsymbol{\Lambda}_{\mu}\boldsymbol{\delta}_{\mu}+2\boldsymbol{\Pi}\boldsymbol{\Lambda}_{\mu}^H\boldsymbol{\Upsilon}\boldsymbol{\Lambda}_{\nu}\boldsymbol{\delta}_{\nu}\notag\\
            &=\boldsymbol{\Pi}M_l(\tB_{\rm c},\tC_{\rm c},\tD).\notag
    \end{align}
    If $M_l(\tB_{\rm c},\tC_{\rm c},\tD)={\bf 0}$, then $M_l(\pi_K\circ_1\tB_{\rm c},\pi_K\circ_{[1,2]}\tC_{\rm c},\pi_K\circ_1\tD)={\bf 0}$ also holds. Therefore, Equation \eqref{eq:kkt1} remains valid under arbitrary permutation along the $K$ dimension. Following the same methodology, we can prove that Equation \eqref{eq:kkt2} holds. Consequently, permutation operations along the $K$ dimension satisfy the KKT conditions and likewise yield optimal solutions. This completes the proof of Equation \eqref{eq:cizf pos TE a}.

    We further consider permutations along the $L$ dimension. Since the optimization problem is solved independently for each $l \in \mathcal{L}$, permuting along the $L$ dimension does not alter the solution at any individual $l$, but merely applies the same permutation to the order of the solutions, which clearly preserves TE. This completes the proof of Equation \eqref{eq:cizf pos TE b}.

    \ifCLASSOPTIONcaptionsoff
    \newpage
    \fi
    
    \bibliographystyle{IEEEtran}
    \bibliography{IEEEfull}

\begin{thebibliography}{10}
\providecommand{\url}[1]{#1}
\csname url@samestyle\endcsname
\providecommand{\newblock}{\relax}
\providecommand{\bibinfo}[2]{#2}
\providecommand{\BIBentrySTDinterwordspacing}{\spaceskip=0pt\relax}
\providecommand{\BIBentryALTinterwordstretchfactor}{4}
\providecommand{\BIBentryALTinterwordspacing}{\spaceskip=\fontdimen2\font plus
\BIBentryALTinterwordstretchfactor\fontdimen3\font minus
  \fontdimen4\font\relax}
\providecommand{\BIBforeignlanguage}[2]{{%
\expandafter\ifx\csname l@#1\endcsname\relax
\typeout{** WARNING: IEEEtran.bst: No hyphenation pattern has been}%
\typeout{** loaded for the language `#1'. Using the pattern for}%
\typeout{** the default language instead.}%
\else
\language=\csname l@#1\endcsname
\fi
#2}}
\providecommand{\BIBdecl}{\relax}
\BIBdecl

\bibitem{lu2014overview}
L.~Lu, G.~Y. Li, A.~L. Swindlehurst, A.~Ashikhmin \emph{et~al.}, ``An overview
  of massive {MIMO}: Benefits and challenges,'' \emph{IEEE J. Sel. Top. Sign.
  Proces.}, vol.~8, no.~5, pp. 742--758, Oct. 2014.

\bibitem{WANG2025}
W.~Wang, Y.~Zhu, Y.~Wang, R.~Ding \emph{et~al.}, ``Toward mobile satellite
  internet: The fundamental limitation of wireless transmission and enabling
  technologies,'' \emph{Engineering}, Early Access, 2025.

\bibitem{1468466}
M.~Joham, W.~Utschick, and J.~Nossek, ``Linear transmit processing in {MIMO}
  communications systems,'' \emph{IEEE Trans. Signal Process.}, vol.~53, no.~8,
  pp. 2700--2712, Aug. 2005.

\bibitem{8359237}
M.~Alodeh, D.~Spano, A.~Kalantari, C.~G. Tsinos \emph{et~al.}, ``Symbol-level
  and multicast precoding for multiuser multiantenna downlink: {A}
  state-of-the-art, classification, and challenges,'' \emph{IEEE Commun. Surv.
  Tutor.}, vol.~20, no.~3, pp. 1733--1757, May 2018.

\bibitem{Li2021}
A.~Li, C.~Masouros, B.~Vucetic, Y.~Li \emph{et~al.}, ``Interference
  exploitation precoding for multi-level modulations: Closed-form solutions,''
  \emph{IEEE Trans. Commun.}, vol.~69, no.~1, pp. 291--308, Jan. 2021.

\bibitem{4801492}
C.~Masouros and E.~Alsusa, ``Dynamic linear precoding for the exploitation of
  known interference in {MIMO} broadcast systems,'' \emph{IEEE Trans. Wireless
  Commun.}, vol.~8, no.~3, pp. 1396--1404, Mar. 2009.

\bibitem{9035662}
A.~Li, D.~Spano, J.~Krivochiza, S.~Domouchtsidis \emph{et~al.}, ``A tutorial on
  interference exploitation via symbol-level precoding: Overview,
  state-of-the-art and future directions,'' vol.~22, no.~2, pp. 796--839, 2nd
  Quart 2020.

\bibitem{5605266}
C.~Masouros, ``Correlation rotation linear precoding for {MIMO} broadcast
  communications,'' \emph{IEEE Trans. Signal Process.}, vol.~59, no.~1, pp.
  252--262, Jan. 2011.

\bibitem{7103338}
C.~Masouros and G.~Zheng, ``Exploiting known interference as green signal power
  for downlink beamforming optimization,'' \emph{IEEE Trans. Signal Process.},
  vol.~63, no.~14, pp. 3628--3640, Jul. 2015.

\bibitem{7417066}
M.~Alodeh, S.~Chatzinotas, and B.~Ottersten, ``Constructive interference
  through symbol level precoding for multi-level modulation,'' in \emph{Proc.
  IEEE Glob. Commun. Conf. (GLOBECOM)}, Dec. 2015, pp. 1--6.

\bibitem{7942010}
------, ``Symbol-level multiuser {MISO} precoding for multi-level adaptive
  modulation,'' \emph{IEEE Trans. Wireless Commun.}, vol.~16, no.~8, pp.
  5511--5524, Aug. 2017.

\bibitem{8575164}
S.~Domouchtsidis, C.~G.~Tsinos, S.~Chatzinotas, and B.~Ottersten,
  ``Symbol-level precoding for low complexity transmitter architectures in
  large-scale antenna array systems,'' \emph{IEEE Trans. Wireless Commun.},
  vol.~18, no.~2, pp. 852--863, Feb. 2019.

\bibitem{8466792}
A.~Li and C.~Masouros, ``Interference exploitation precoding made practical:
  Optimal closed-form solutions for {PSK} modulations,'' \emph{IEEE Trans.
  Wireless Commun.}, vol.~17, no.~11, pp. 7661--7676, Nov. 2018.

\bibitem{8299553}
A.~Haqiqatnejad, F.~Kayhan, and B.~Ottersten, ``Constructive interference for
  generic constellations,'' \emph{IEEE Signal Process. Lett.}, vol.~25, no.~4,
  pp. 586--590, Apr. 2018.

\bibitem{8477154}
------, ``Symbol-level precoding design based on distance preserving
  constructive interference regions,'' \emph{IEEE Trans. Signal Process.},
  vol.~66, no.~22, pp. 5817--5832, Nov. 2018.

\bibitem{8815429}
------, ``An approximate solution for symbol-level multiuser precoding using
  support recovery,'' in \emph{Proc. IEEE Workshop Signal Process. Adv.
  Wireless Commun. (SPAWC)}, Jul. 2019, pp. 1--5.

\bibitem{wang2024symbol}
Y.~Wang, H.~Hou, W.~Wang, and X.~Yi, ``Symbol-level precoding for average {SER}
  minimization in multiuser {MISO} systems,'' \emph{IEEE Wireless Commun.
  Lett.}, vol.~13, no.~4, pp. 1103--1107, Apr. 2024.

\bibitem{shao2020minimum}
M.~Shao, Q.~Li, and W.-K. Ma, ``Minimum symbol-error probability symbol-level
  precoding with intelligent reflecting surface,'' \emph{IEEE Wireless Commun.
  Lett.}, vol.~9, no.~10, pp. 1601--1605, Oct. 2020.

\bibitem{mohammad2021unsupervised}
A.~Mohammad, C.~Masouros, and Y.~Andreopoulos, ``An unsupervised deep unfolding
  framework for robust symbol-level precoding,'' \emph{IEEE Open J. Commun.
  Soc.}, vol.~4, pp. 1075--1090, Apr. 2023.

\bibitem{8647896}
A.~Haqiqatnejad, F.~Kayhan, and B.~Ottersten, ``Robust design of power
  minimizing symbol-level precoder under channel uncertainty,'' in \emph{Proc.
  IEEE Glob. Commun. Conf. (GLOBECOM)}, Dec. 2018, pp. 1--6.

\bibitem{hegde2019interference}
G.~Hegde, C.~Masouros, and M.~Pesavento, ``Interference exploitation-based
  hybrid precoding with robustness against phase errors,'' \emph{IEEE Trans.
  Wireless Commun.}, vol.~18, no.~7, pp. 3683--3696, Jul. 2019.

\bibitem{9025051}
A.~Haqiqatnejad, F.~Kayhan, and B.~Ottersten, ``Robust {SINR}-constrained
  symbol-level multiuser precoding with imperfect channel knowledge,''
  \emph{IEEE Trans. Signal Process.}, vol.~68, pp. 1837--1852, Mar. 2020.

\bibitem{7042789}
M.~Alodeh, S.~Chatzinotas, and B.~Ottersten, ``Constructive multiuser
  interference in symbol level precoding for the {MISO} downlink channel,''
  \emph{IEEE Trans. Signal Process.}, vol.~63, no.~9, pp. 2239--2252, May 2015.

\bibitem{7472286}
K.~L. Law and C.~Masouros, ``Constructive interference exploitation for
  downlink beamforming based on noise robustness and outage probability,'' in
  \emph{IEEE Int. Conf. Acoust., Speech Signal Process. (ICASSP)}, Mar. 2016,
  pp. 3291--3295.

\bibitem{8374931}
------, ``Symbol error rate minimization precoding for interference
  exploitation,'' \emph{IEEE Trans. Commun.}, vol.~66, no.~11, pp. 5718--5731,
  Nov. 2018.

\bibitem{9910472}
Y.~Wang, W.~Wang, L.~You, C.~G. Tsinos \emph{et~al.}, ``Weighted {MMSE}
  precoding for constructive interference region,'' \emph{IEEE Wireless Commun.
  Lett.}, vol.~11, no.~12, pp. 2605--2609, Dec. 2022.

\bibitem{9416909}
J.~Zheng, J.~Zhang, E.~Björnson, and B.~Ai, ``Impact of channel aging on
  cell-free massive {MIMO} over spatially correlated channels,'' \emph{IEEE
  Trans. Wireless Commun.}, vol.~20, no.~10, pp. 6451--6466, Oct. 2021.

\bibitem{hou2024tensor}
H.~Hou, Y.~Wang, Y.~Zhu, X.~Yi \emph{et~al.}, ``A tensor-structured approach to
  dynamic channel prediction for massive {MIMO} systems with temporal
  non-stationarity,'' \emph{arXiv preprint arXiv:2412.06713}, 2024.

\bibitem{10804143}
J.~Zhuang, X.~He, Y.~Wang, J.~Liu \emph{et~al.}, ``Covnet: Covariance
  information-assisted {CSI} feedback for {FDD} massive {MIMO} systems,''
  \emph{IEEE Wireless Commun. Lett.}, vol.~14, no.~3, pp. 641--645, Mar. 2025.

\bibitem{hou2024joint}
H.~Hou, Y.~Wang, X.~Yi, W.~Wang \emph{et~al.}, ``Joint beam alignment and
  doppler estimation for fast time-varying wideband mmwave channels,''
  \emph{IEEE Trans. Wireless Commun.}, vol.~23, no.~9, pp. 10\,895--10\,910,
  Sept. 2024.

\bibitem{zhuang2025extract}
J.~Zhuang, Y.~Wang, H.~Hou, Y.~Han \emph{et~al.}, ``Extract the best, discard
  the rest: {CSI} feedback with offline large {AI} models,'' \emph{arXiv
  preprint arXiv:2505.08566}, 2025.

\bibitem{8694866}
A.-A. Lu, X.~Gao, W.~Zhong, C.~Xiao \emph{et~al.}, ``Robust transmission for
  massive {MIMO} downlink with imperfect {CSI},'' \emph{IEEE Trans. Commun.},
  vol.~67, no.~8, pp. 5362--5376, Aug. 2019.

\bibitem{confPaper}
Y.~Wang, X.~Yi, H.~Hou, and W.~Wang, ``Robust symbol-level precoding for {MIMO}
  downlink transmission with channel aging,'' in \emph{IEEE Glob. Commun. Conf.
  (GLOBECOM)}, Dec. 2023, pp. 5763--5768.

\bibitem{wang2024robust}
Y.~Wang, X.~Yi, H.~Hou, W.~Wang \emph{et~al.}, ``Robust symbol-level precoding
  for massive {MIMO} communication under channel aging,'' \emph{IEEE Trans.
  Wireless Commun.}, vol.~23, no.~9, pp. 10\,864--10\,878, Sept. 2024.

\bibitem{9685567}
A.~Mohammad, C.~Masouros, and Y.~Andreopoulos, ``An unsupervised learning-based
  approach for symbol-level-precoding,'' in \emph{Proc. IEEE Glob. Commun.
  Conf. (GLOBECOM)}, Dec. 2021, pp. 1--6.

\bibitem{8465957}
A.~Haqiqatnejad, F.~Kayhan, and B.~Ottersten, ``Power minimizer symbol-level
  precoding: A closed-form suboptimal solution,'' \emph{IEEE Signal Process.
  Lett.}, vol.~25, no.~11, pp. 1730--1734, Nov. 2018.

\bibitem{8647428}
J.~Krivochiza, J.~Merlano-Duncan, S.~Andrenacci, S.~Chatzinotas \emph{et~al.},
  ``Closed-form solution for computationally efficient symbol-level
  precoding,'' in \emph{Proc. IEEE Glob. Commun. Conf. (GLOBECOM)}, Dec. 2018,
  pp. 1--6.

\bibitem{11049893}
Y.~Wang, H.~Hou, X.~Yi, W.~Wang \emph{et~al.}, ``Towards unified {AI} models
  for {MU-MIMO} communications: A tensor equivariance framework,'' \emph{IEEE
  Trans. Wireless Commun.}, Early Access, 2025.

\bibitem{wang2024soft}
Y.~Wang, H.~Hou, W.~Wang, X.~Yi \emph{et~al.}, ``Soft demodulator for
  symbol-level precoding in coded multiuser {MISO} systems,'' \emph{IEEE Trans.
  Wireless Commun.}, vol.~23, no.~10, pp. 14\,819--14\,835, Oct. 2024.

\bibitem{pratik2020re}
K.~Pratik, B.~D. Rao, and M.~Welling, ``{RE-MIMO}: Recurrent and permutation
  equivariant neural {MIMO} detection,'' \emph{IEEE Trans. Signal Process.},
  vol.~69, pp. 459--473, Dec. 2020.

\bibitem{wang2025statistical}
Y.~Wang, V.~N. Ha, K.~Ntontin, H.~Yan \emph{et~al.}, ``Statistical {CSI}-based
  distributed precoding design for {OFDM}-cooperative multi-satellite
  systems,'' \emph{arXiv preprint arXiv:2505.08038}, 2025.

\bibitem{8602458}
Y.~Choi, J.~Lee, M.~Rim, and C.~G. Kang, ``Constructive interference
  optimization for data-aided precoding in multi-user {MISO} systems,''
  \emph{IEEE Trans. Wireless Commun.}, vol.~18, no.~2, pp. 1128--1141, Jan.
  2019.

\bibitem{li2020symbol}
A.~Li, F.~Liu, X.~Liao, Y.~Shen \emph{et~al.}, ``Symbol-level precoding made
  practical for multi-level modulations via block-level rescaling,'' in
  \emph{IEEE Workshop Signal Process. Adv. Wireless Commun. (SPAWC)}, Sept.
  2021, pp. 71--75.

\bibitem{lawson1995solving}
C.~L. Lawson and R.~J. Hanson, \emph{Solving least squares problems}.\hskip 1em
  plus 0.5em minus 0.4em\relax Philadelphia, PA, USA: SIAM, 1995.

\bibitem{zaheer2017deep}
M.~Zaheer, S.~Kottur, S.~Ravanbakhsh, B.~Poczos \emph{et~al.}, ``Deep sets,''
  \emph{Neural Inf. Proces. Syst. (NeurIPS)}, vol.~30, Dec. 2017.

\bibitem{hartford2018deep}
J.~Hartford, D.~Graham, K.~Leyton-Brown, and S.~Ravanbakhsh, ``Deep models of
  interactions across sets,'' in \emph{Int. Conf. Mach. Learn. (ICML)}, Jul.
  2018, pp. 1909--1918.

\bibitem{yun2019transformers}
C.~Yun, S.~Bhojanapalli, A.~S. Rawat, S.~J. Reddi \emph{et~al.}, ``Are
  transformers universal approximators of sequence-to-sequence functions?'' in
  \emph{Int. Conf. Learn. Represent. (ICLR)}, Apr. 2019.

\bibitem{kim2022pure}
J.~Kim, D.~Nguyen, S.~Min, S.~Cho \emph{et~al.}, ``Pure transformers are
  powerful graph learners,'' \emph{Neural Inf. Proces. Syst. (NeurIPS)},
  vol.~35, pp. 14\,582--14\,595, Nov. 2022.

\bibitem{lee2019set}
J.~Lee, Y.~Lee, J.~Kim, A.~Kosiorek \emph{et~al.}, ``Set transformer: A
  framework for attention-based permutation-invariant neural networks,'' in
  \emph{Int. Conf. Mach. Learn. (ICML)}, Jun. 2019, pp. 3744--3753.

\bibitem{keriven2019universal}
N.~Keriven and G.~Peyr{\'e}, ``Universal invariant and equivariant graph neural
  networks,'' \emph{Neural Inf. Proces. Syst. (NeurIPS)}, vol.~32, Dec. 2019.

\bibitem{pan2022permutation}
H.~Pan and R.~Kondor, ``Permutation equivariant layers for higher order
  interactions,'' in \emph{Proc. Int. Conf. Artif. Intell. Statist. (AISTATS)},
  Mar. 2022, pp. 5987--6001.

\bibitem{Woo_2018_ECCV}
S.~Woo, J.~Park, J.-Y. Lee, and I.~S. Kweon, ``{CBAM}: Convolutional block
  attention module,'' in \emph{Proc. Eur. Conf. Comput. Vis. (ECCV)}, September
  2018.

\bibitem{He_2016_CVPR}
K.~He, X.~Zhang, S.~Ren, and J.~Sun, ``Deep residual learning for image
  recognition,'' in \emph{Proc. IEEE Conf. Comput. Vis. Pattern Recog. (CVPR)},
  June 2016.

\bibitem{hou2025tensor}
H.~Hou, Y.~Wang, X.~Yi, W.~Wang \emph{et~al.}, ``Tensor-structured bayesian
  channel prediction for upper mid-band {XL-MIMO} systems,'' \emph{arXiv
  preprint arXiv:2508.08491}, 2025.

\bibitem{6758357}
S.~Jaeckel, L.~Raschkowski, K.~Börner, and L.~Thiele, ``Quadriga: A {3-D}
  multi-cell channel model with time evolution for enabling virtual field
  trials,'' \emph{IEEE Trans. Antennas Propag.}, vol.~62, no.~6, pp.
  3242--3256, Mar. 2014.

\bibitem{3gpp_tr_38901_v190}
3GPP, ``Tr 38.901 v19.0.0: Study on channel model for frequencies from 0.5 to
  100 {GH}z,'' 3GPP, Tech. Rep. TR 38.901 V19.0.0, Jun 2025.

\bibitem{1391204}
C.~Peel, B.~Hochwald, and A.~Swindlehurst, ``A vector-perturbation technique
  for near-capacity multiantenna multiuser communication-part {I}: {C}hannel
  inversion and regularization,'' \emph{IEEE Trans. Commun.}, vol.~53, no.~1,
  pp. 195--202, Jan. 2005.

\end{thebibliography}

\end{document}